\documentclass[letterpaper,aps,twocolumn,showpacs,superscriptaddress,floats,prb]{revtex4}
 
\usepackage{amsmath,color}
\usepackage{graphicx,epsfig,dsfont}
\usepackage{amssymb}

\newcommand{\nn}{\nonumber}

\begin{document}


\def\Xint#1{\mathchoice
   {\XXint\displaystyle\textstyle{#1}}%
   {\XXint\textstyle\scriptstyle{#1}}%
   {\XXint\scriptstyle\scriptscriptstyle{#1}}%
   {\XXint\scriptscriptstyle\scriptscriptstyle{#1}}%
   \!\int}
\def\XXint#1#2#3{{\setbox0=\hbox{$#1{#2#3}{\int}$}
     \vcenter{\hbox{$#2#3$}}\kern-.5\wd0}}
\def\ddashint{\Xint=}
\def\dashint{\Xint-}
 

\title{Multiscale quantum criticality: Pomeranchuk instability in isotropic metals}

\author{Mario Zacharias}
\affiliation{Institut f\"ur Theoretische Physik, Universit\"at zu K\"oln,
Z\"ulpicher Str. 77, 50937 K\"oln, Germany
}
\author{Peter W\"olfle}
\affiliation{Institut f\"ur Theorie der Kondensierten Materie, Universit\"at Karlsruhe, 76128 Karlsruhe, Germany}
\author{Markus Garst}
\affiliation{Institut f\"ur Theoretische Physik, Universit\"at zu K\"oln,
Z\"ulpicher Str. 77, 50937 K\"oln, Germany
}

\begin{abstract}
As a paradigmatic example of multiscale quantum criticality, we consider the Pomeranchuk instability of an isotropic Fermi liquid in two spatial dimensions, $d=2$. The corresponding Ginzburg-Landau theory for the quadrupolar fluctuations of the Fermi surface consists of two coupled modes, critical at the same point, and characterized by different dynamical exponents: one being ballistic with dynamical exponent $z=2$ and the other one is Landau damped with $z=3$, thus giving rise to multiple dynamical scales. We find that at temperature $T=0$, the ballistic mode governs the low-energy structure of the theory as it possesses the smaller effective dimension $d+z$. Its self-interaction leads to logarithmic singularities, which we treat with the help of the renormalization group. At finite temperature, the coexistence of two different dynamical scales gives rise to a modified quantum-to-classical crossover. It extends over a parametrically large regime with intricate interactions of quantum and classical fluctuations leading to a universal $T$ dependence of the correlation length independent of the interaction amplitude. The multiple scales are also reflected in the phase diagram and in the critical thermodynamics. In particular, we find that the latter cannot be interpreted in terms of only a single dynamical exponent whereas, e.g., the critical specific heat is determined by the $z=3$ mode, the critical compressibility is found to be dominated by the $z=2$ fluctuations.  
\end{abstract}

\date{\today}

\maketitle

\section{Introduction}

At a quantum phase transition, statics and dynamics of critical fluctuations are inseparably entangled. This is manifest in the dependence of static and thermodynamic properties on the dynamical exponent $z$, that  allows to associate a thermal length, $\xi_T \sim T^{-1/z}$, with temperature $T$. This thermal length is at the origin of the quantum-to-classical crossover where critical fluctuations change their character.\cite{Sachdev,ZinnJustin} Whereas critical bosonic modes with momenta higher than the inverse thermal length are of a quantum-mechanical nature, modes with smaller momenta behave effectively classical. This crossover also reveals itself in the phase diagram spanned by the control parameter of the quantum phase transition, $r$, and temperature, $T$. The behavior of, e.g., thermodynamic properties changes qualitatively when the correlation length $\xi$ is of the same order as the thermal length $\xi_T$.
   
The identification of the critical fluctuations and their dynamics close to a putative quantum phase transition in a material is a nontrivial task. Frequently, low-energy fluctuations characterized by different dynamics coexist and interact. In fact, this is generally the case in critical metals where the dynamics of the order parameter coexists with the dynamics of the fermionic quasiparticles. The standard Hertz-Millis-Moriya model\cite{Hertz76,Moriya95,Millis93,Loehneysen07} for a magnetic instability in a metal assumes that only the dynamics of the paramagnons is important for the characterization of critical properties. Although this assumption has proved to be successful in interpreting a number of experiments, it has failed to explain consistently the properties of the heavy fermion compounds CeCu$_{6-x}$Au$_x$ and YbRh$_2$Si$_2$.\cite{Loehneysen07,Gegenwart08} In particular, it has been argued that the latter system is characterized by the presence of multiple scales,\cite{Gegenwart07} that might be due to the presence of coexisting dynamics.
Theoretically, it is now established that the assumptions of the Hertz-Millis-Moriya approach fail in the case of the ferromagnetic instability.\cite{Loehneysen07} The  coupling of the ferromagnetic fluctuations to the fermionic degrees of freedom gives rise to singular corrections that prevent the formulation of an effective theory exclusively in terms of order-parameter fluctuations.\cite{Vojta97,Belitz00,Belitz04,Rech06}  The presence of multiple dynamics is also important for the antiferromagnetic quantum phase transition in metals in spatial dimension $d=2$; the fermions mediate an effectively long-ranged interaction among the paramagnons resulting in a strongly coupled theory with multiple dynamical scales.\cite{Abanov04}

Although quantum criticality with multiple dynamical scales seems to be rather common, systematic investigations are rare and the generic features of this problem are therefore only poorly understood. The presence of coexisting dynamics and, consequently, different dynamical exponents complicates the analysis of experimental data considerably, as simple scaling relations are bound to fail. Generally speaking, different physical quantities might be dominated by one or the other dynamics such that the interpretation in terms of a single exponent $z$ results in apparent inconsistencies. 
On the theoretical side, a critical field theory with multiple exponents $z$ promises to exhibit a rich structure. The modes with the larger dynamical exponent $z_>$ have a larger phase space available and are thus expected to dominate, e.g., the specific heat. The modes with the smaller exponent $z_<$, on the other hand, have a smaller effective dimension $d+z_<$ and might therefore trigger infrared singularities in perturbative loop corrections. In addition, multiple dynamical exponents imply multiple thermal scales $\xi_T$ with the concomitant crossovers. Instead of a single quantum-to-classical crossover one expects an extended crossover regime with coexisting and interacting quantum and classical fluctuations.  

In the present work, we theoretically investigate these phenomena in a simple model exhibiting multiscale quantum criticality.
Our aim here is not to go beyond the Hertz-Millis-Moriya framework. Instead, we still limit ourselves to an effective bosonic theory where the fluctuations are however characterized by multiple dynamical scales. To be specific, we consider the effective Ginzburg-Landau theory for the Pomeranchuk instability\cite{Pomeranchuk54} in an isotropic metal as proposed by Oganesyan {\it et al}.\cite{Oganesyan01} The order parameter is given by the shear modes of the Fermi sphere that drive a spontaneous deformation of the Fermi surface at the instability resulting in a ``nematic'' state. In $d=2$, there are two critical bosonic shear modes characterized by different dynamical exponents. Whereas one is damped by particle-hole excitations in the metal and has a dynamical exponent $z_>=3$, the other is ballistic and undamped with $z_<=2$. We confine ourselves to two spatial dimension, $d=2$, where symmetry consideration allow the Pomeranchuk transition to be of second order and fluctuation corrections are pronounced.

The Pomeranchuk instability and its criticality has received renewed interest in recent years. 
In analogy to paramagnons close to a magnetic quantum phase transition in metals, the critical collective shear fluctuations close to a Pomeranchuk quantum critical point give rise to a singular interaction among the fermionic degrees of freedom leading to a breakdown of Fermi-liquid theory.\cite{Oganesyan01,Metzner03,DellAnna06} The structure of the critical theory, however, depends on the presence of an underlying crystal lattice. Whereas in an isotropic metal the Fermi-liquid properties are destroyed over the full Fermi surface,\cite{Oganesyan01} well-defined quasiparticles partly survive if a crystal lattice breaks rotational symmetry from the outset.\cite{Metzner03,DellAnna06} Moreover, the aforementioned undamped $z_< = 2$ mode only becomes critical in the isotropic case but is always gapped if a crystal is present. A pecularity of the nematic, i.e, Pomeranchuk-ordered phase in an isotropic medium is the existence of a Landau-damped Goldstone mode that also leads to singular Fermi-liquid corrections thus providing a rare example of an extended non-Fermi-liquid phase.\cite{Oganesyan01} An extensive investigation of the mean-field properties of the Pomeranchuk transition has been carried out by Yamase and collaborators.\cite{Yamase04,Yamase05,Yamase07,Yamase09} A nonperturbative analysis of Pomeranchuk quantum criticality was presented in Refs.~\onlinecite{Lawler06,Chubukov06,Lawler07}, and a discussion of the Pomeranchuk instability from the perspective of Fermi-liquid theory was given by one of the authors.\cite{Woelfle07} Quintanilla {\it et al.}\cite{Quintanilla06,Quintanilla08} presented a careful study of as to how the instability might arise from central interactions among quasiparticles. Finally, the influence of disorder on the transition was studied in Ref.~\onlinecite{Ho08}. 
Nematic ordering was also discussed extensively in the context of cuprate superconductors, for a recent review see Ref.~\onlinecite{Vojta09}. On the experimental side, a metallic nematic phase in Sr$_3$Ru$_2$O$_7$ close to its metamagnetic transition has been reported by Borzi {\it et al.}\cite{Borzi07,Fradkin07} A Pomeranchuk instability in an isotropic metal, that we consider here, has not been realized so far but might be achievable in cold atom systems close to a Feshbach resonance in a channel with higher angular momentum.

\subsection{Summary of main results}
\label{sec:SummaryResults}

\begin{figure}
\includegraphics[width=0.4\textwidth]{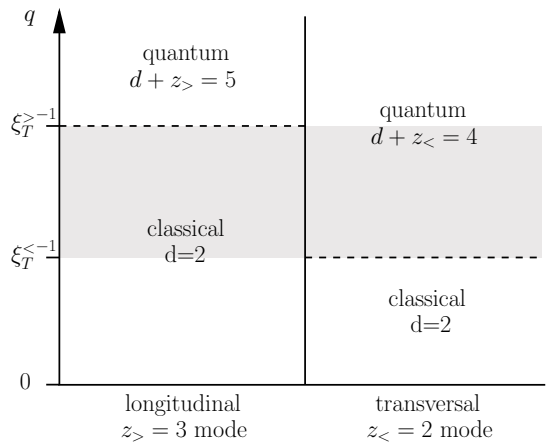} 
\caption{Effective dimensions of the longitudinal and transversal shear fluctuations as a function of momentum $q$. The respective thermal momenta, $\xi_T^{-1} \sim T^{1/z}$, separate two regimes where the fluctuations have either a \hspace*{.1em}quantum \hspace*{.1em}character, \hspace*{.1em}$q > \xi_T^{-1}$, with effective dimension $d+z$ or a classical character, $q < \xi_T^{-1}$, with effective dimension $d=2$. There is an extended quantum-to-classical crossover (gray shaded), where the quantum regime of the transversal $z_< = 2$ mode overlaps with the classical regime of the longitudinal mode $z_> = 3$.
}
\label{fig:ModeMomenta}
\end{figure}

In the literature, attention has been mainly focused on the damped $z_>=3$ component of the shear fluctuations. Oganesyan {\it et al.}\cite{Oganesyan01} even argued that the undamped $z_< = 2$ fluctuation mode plays no role in the critical theory because its dynamics seems to be irrelevant with respect to $z_>=3$ scaling; hence, the critical theory should be fully captured by a Gaussian theory as the effective dimension $d+z_> = 5$ is larger than the upper critical dimension of the effective $\Phi^4$ theory, $d^+_c = 4$. In contrast to this, we find that the scaling arguments of Ref.~\onlinecite{Oganesyan01} are misleading and that the $z_< = 2$ mode is in fact instrumental for the Pomeranchuk quantum phase transition in $d=2$. 
%
%
The ballistic mode has an effective dimension $d+z_< = 4 = d^+_c$, and it therefore generates important logarithmic singularities at $T=0$ in loop corrections. We treat these singularities with the help of the renormalization group (RG). The derived RG equations identify the universality class which bears signatures of the tensorial nature of the shear fluctuations and differs, in particular, from the ones with Ising as well as XY symmetry. 
By solving the RG equation, we find, for example, that the correlation length $\xi$ diverges with vanishing distance to the quantum critical point, $r \to 0^+$, according to
\begin{align} \label{Gap-1}
\left.\xi^{-2}\right|_{T=0} \sim \frac{r}{(\log\frac{\bar{\Lambda}^2}{r})^{4/9}},
\end{align}
where $\bar{\Lambda}$ is some momentum cutoff. The characteristic logarithmic enhancement with exponent $4/9$ explicitly demonstrates that interactions among the $z_<=2$ modes lead to qualitative corrections that are beyond Gaussian critical behavior.

At any finite temperatures, the coexisting dynamics implies the presence of two thermal lengths, $\xi^>_T \sim T^{-1/z_>}$ and $\xi^<_T \sim T^{-1/z_<}$, that give rise to an interesting modification of the quantum-to-classical crossover. 
Generally at criticality, $r=0$, critical modes change their character at the momentum scale of the inverse thermal length, $q \sim \xi_T^{-1}$.  For larger momenta $q > \xi_T^{-1}$, the fluctuations are essentially of a quantum-mechanical nature because the associated Matsubara frequencies can be approximated to be dense. For smaller momenta $q < \xi_T^{-1}$, on the other hand, fluctuations are effectively classical in the sense that they are dominated by the Matsu\-bara zero mode. In the presence of only a single thermal length $\xi_T$, an effective theory at criticality, $r=0$, can be formulated that changes its character at the momentum scale $q \sim \xi^{-1}_T$ from a quantum theory with dimension $d + z$ to an effective classical theory with dimension $d$, which is also known as dimensional reduction.\cite{Sachdev, ZinnJustin,Millis93,Sachdev97} In the present problem of the Pomeranchuk instability we find, however, that it is not possible to derive the critical properties from an effective classical model so that the concept of dimensional reduction breaks down. 
As illustrated in Fig.~\ref{fig:ModeMomenta}, the two thermal lengths give rise to an extended intermediate momentum regime where the $z_< = 2$ fluctuation mode is still quantum and the $z_> = 3$ mode is already classical. Remarkably, it turns out that interactions between quantum and classical fluctuations with momenta belonging to this extended quantum-to-classical crossover regime determine the critical properties at $r=0$ resulting in the inapplicability of dimensional reduction.
At $d+ z_< = 4 = d^+_c$, the $z_<=2$ quantum fluctuations generate scale-dependent vertices in loop corrections, that are in turn probed by the classical $z_>=3$ mode. 
This peculiar interplay of quantum and classical fluctuations yields, for example, a universal temperature dependence of the correlation length in the low-$T$ limit,
\begin{align} \label{CorrLengthUniversalT}
\left.\xi^{-2}\right|_{r=0} = \mathcal{C}\,  T,\qquad \mathcal{C} = 4 \left[1 - \left(\frac{2}{3}\right)^{4/9}\right],
\end{align} 
where $\xi$ and $T$ are measured in dimensionless units to be specified below. The $T$ dependence is universal in the sense that it does not depend on the value of the $\Phi^4$-interaction amplitude although it is triggered by the very same interaction. Note that the correlation length [Eq.~(\ref{CorrLengthUniversalT})] \hspace*{.1em}is \hspace*{.1em}of \hspace*{.1em}the \hspace*{.1em}same \hspace*{.1em}order \hspace*{.1em}as \hspace*{.1em}the \hspace*{.1em}thermal \hspace*{.1em}length \hspace*{.2em}$\xi_T^< \sim T^{-1/2}$ which is a clear manifestation for the breakdown of dimensional reduction.
 
\begin{figure}
\includegraphics[width=0.45\textwidth]{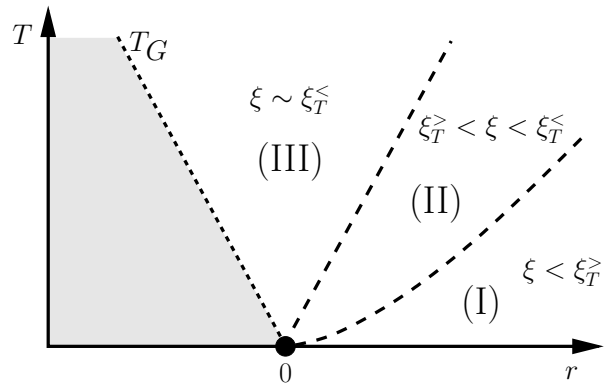} 
\caption{Phase diagram for the Pomeranchuk quantum phase transition in $d=2$. The correlation length $\xi$ and the two thermal lengths, $\xi^>_T \sim T^{-1/3}$ and $\xi^{<}_T \sim T^{-1/2}$, distinguish three regimes: (I) a Fermi-liquid regime where $\xi$ is the smallest scale, (II) an overlap regime where $\xi$ is sandwiched between the two thermal lengths, and (III) a quantum critical regime where $\xi\sim \xi^>_T$ is the largest scale. The line labeled $T_G(r)$ represents the Ginzburg temperature. }
\label{fig:PD}
\end{figure}

The coexistence of two dynamical scales is also reflected in the phase diagram and in the critical thermodynamics. The two thermal lengths $\xi_T^> < \xi_T^<$ 
divide the disordered region of the phase diagrams into three regimes, see Fig.~\ref{fig:PD}: (I) a low-temperature regime for $\xi < \xi_T^>$ , (II) an overlap regime, $\xi_T^>  <\xi <\xi_T^<$, and (III) a quantum critical regime, $\xi_T^> \sim \xi$. Neglecting logarithmic corrections, the two crossovers are located in the phase diagram at $T \sim r^{3/2}$ and $T \sim r$, respectively. In the present work, we limit ourselves to the disordered side of the phase diagram on the right-hand side of the Ginzburg temperature, $T_G(r)$ (non-shaded in Fig.~\ref{fig:PD}).

It is {\it a priori} not clear which of the two crossover scales is more important for thermodynamics. Our calculations reveal that it depends in fact on the particular thermodynamic quantity. 
Due to their larger phase space, the $z_> = 3$ fluctuations dominate the specific heat coefficient $\gamma$. As a consequence, $\gamma$ changes its behavior at the (I)/(II) crossover, $\xi \sim \xi_T^>$. 
The thermal expansion, $\alpha$, is at first sight also dominated by the $z_> =3$ mode. However, the repeated scattering of $z_< = 2$ quantum fluctuations leads to a logarithmic dependence of the thermal expansion vertex. This in turn results in a sensitivity of the thermal expansion on the (II)/(III) crossover too, and $\alpha$ therefore changes its critical behavior at both crossovers. 
Finally, the asymptotic critical behavior of the compressibility, $\kappa$, is dominated by the $z_< = 2$ mode and it is thus sensitive to the (II)/(III) crossover only. 
To summarize, the critical thermodynamics knows about both crossovers, (I)/(II) and (II)/(III) in Fig.~\ref{fig:PD} and cannot be interpreted in terms of only a single dynamical exponent.
 
\subsection{Outline of the paper}

In Sec.~\ref{sec:PomFL}, we review the derivation of the effective Ginzburg-Landau theory of the Pomeranchuk instability in an isotropic metal of Ref.~\onlinecite{Oganesyan01}. In addition, we discuss corrections to the shear susceptibility beyond the random-phase approximation (RPA), and we point out the importance of vertex corrections in order to maintain the $z_< = 2$ dynamics of the undamped shear mode. Readers mostly interested in the main results might directly proceed to Sec.~\ref{sec:BosonicTheory}, where the effective bosonic field theory is analyzed, and the critical thermodynamics is discussed. We end with a summary of the results in Sec.~\ref{sec:discussion}. Detailed calculations are relegated to the appendices.

\section{Pomeranchuk instability of the Fermi liquid}
\label{sec:PomFL}

We shortly review the derivation of the effective Ginz-burg-Landau theory for the Pomeranchuk instability of an isotropic metal by closely following Oganesyan {\it et al}.\cite{Oganesyan01} We limit ourselves to considerations in spatial dimensions $d=2$. The starting point is a model of spinless fermions interacting via a quadrupolar interaction,
\begin{align} \label{model}
\mathcal{S} = \int d\tau d^d x \left[
\Psi^\dag g_0^{-1} \Psi  + \frac{F_2}{2}  \left(\Psi^\dag \hat{Q}_{ij} \Psi\right) \left(\Psi^\dag \hat{Q}_{ji} \Psi\right) \right],
\end{align}
where $g_0^{-1}(k,i\omega_n) = -i\omega_n + \varepsilon_k - \mu$ is the fermionic Green's function.
The operator $\hat{Q}_{ij}$ is defined by its Fourier transform with respect to the spatial coordinate, $\Psi_{\bf x}^\dag \hat{Q}_{ij} \Psi_{{\bf x'}} \to \Psi_{\bf k}^\dag Q_{ij}[({\bf k}+{\bf k'})/2] \Psi_{\bf k'}$ where the quadrupolar momentum tensor is
\begin{align}
Q_{ij}({\bf k}) &=  d\, \hat{\bf k}_i \hat{\bf k}_j - \delta_{ij},
\end{align}  
with $d=2$, $\hat{\bf k}_i = {\bf k}_i/k$ and $i,j \in \{1,2\}$.
%
%
%
The expectation value $\varepsilon_{ij} = \langle \Psi^\dag \hat{Q}_{ij} \Psi \rangle$
is the quadrupole density and can be interpreted as the traceless part of a strain tensor representing the elastic shear modes of the Fermi sphere. With the help of the Hubbard-Stratonovich transformation defined by the action
\begin{align}
\mathcal{S} = \int d\tau d^d x \left[
\Psi^\dag g_0^{-1} \Psi  - \frac{1}{2 F_2} \sigma_{ij} \sigma_{ij} +  \sigma_{ij} \left(\Psi^\dag \hat{Q}_{ij} \Psi\right) \right],
\end{align}
we can introduce the quadrupolar stress tensor $\sigma_{ij}$ that is real, symmetric, and traceless. Upon integrating out the fermionic degrees of freedom one obtains an effective elasticity theory for the Fermi liquid in terms of $\sigma_{ij}$. 

General symmetry considerations require that its static Ginzburg-Landau energy functional has the form
\begin{align} \label{EnergyFunctional}
\mathcal{E}[\sigma] = 
{\rm tr } \left\{ 
\frac{r_0}{2} \sigma^2 - \frac{K}{2} \sigma \nabla^2 \sigma  - 
\frac{K_2}{12} \sigma^2 D \sigma +
\frac{u}{12} \sigma^4 + \dots \right\}
\end{align}
with the quadrupolar gradient tensor $D_{ij} \!=\! d\, \partial_{i} \partial_{j} \!-\! \delta_{ij}\! \nabla^2$.
Note that generally in $d=2$ a trace over a product of three quadrupolar tensors vanishes. In particular, a term tr$\{\sigma D \sigma\}$ is absent so that there is only a single Frank constant $K$ on the quadratic level of the theory in the lowest-order gradient expansion. In the following, we will neglect the term with constant $K_2$ as it is irrelevant for our considerations. We note in passing that the analyticity of the gradient expansion employed in Eq.~(\ref{EnergyFunctional}) is robust against interaction effects; it was argued in Ref.~\onlinecite{Rech06} that the nonanalytic momentum dependences that are generated close to a ferromagnetic quantum critical point are absent for the Pomeranchuk instability. 

The quadrupolar gradient tensor also allows for a coupling of the shear modes to the compressional mode, i.e., the isotropic density fluctuation, $\delta n$, in the Fermi liquid of the form $\delta n\, {\rm tr}\{D \sigma\}$. If this coupling is strong, it hybridizes the two different modes giving rise to low-energy excitations at some finite momentum. We will assume that this coupling is sufficiently weak such that we can neglect the compressional fluctuations in the following.

The dynamics of the shear modes $\sigma_{ij}$ are derived from an explicit calculation of the quadrupolar polarization of the Fermi liquid.\cite{Oganesyan01} For this purpose, it is convenient to expand the shear field at a given momentum ${\bf q}$ and Matsubara frequency $\Omega_n$, $\sigma_{ij}({\bf q}, i\Omega_n)$, into components longitudinal and transversal to the quadrupolar momentum tensor,
\begin{align}
\sigma_{ij}({\bf q}, i\Omega_n) =  
\phi_{\parallel {\bf q}, i\Omega_n} E^{\parallel}_{ij}(\hat{\bf q}) + \phi_{\perp {\bf q}, i\Omega_n}  E^{\perp}_{ij}(\hat{\bf q}).
\end{align}
The matrices $E^{\parallel,\perp}$ form a basis set for $2\times2$ real, symmetric and traceless tensors and are
defined with respect to the momentum direction $\hat{\bf q} = {\bf q}/q$,
\begin{subequations} \label{Eigenvectors}
\begin{align}
 E^{\parallel}_{ij}(\hat{\bf q}) &= \frac{1}{\sqrt{2}} \left(2 \hat{q}_i \hat{q}_j -  \delta_{ij}  \right) = \frac{1}{\sqrt{2}}  \left(\hat{q}_i \hat{q}_j - \hat{p}_i \hat{p}_j \right),
\\
 E^{\perp}_{ij}(\hat{\bf q}) &=  \frac{1}{\sqrt{2}}  \left(  \hat{q}_i \hat{p}_j + \hat{p}_i \hat{q}_j  \right),
\end{align}
\end{subequations}
where the unit vectors $\hat{\bf p}$ and $\hat{\bf q}$ are orthogonal with det$(\hat{\bf p},\hat{\bf q})=1$.
The basis set is normalized $E^\alpha_{ij}E^\beta_{ij} = \delta_{\alpha \beta}$, with $\alpha, \beta \in \{\parallel, \perp\}$. The matrices of different basis sets transform according to
\begin{align} \label{Trafos}
E^{\alpha}_{ij}(\hat{\bf k}) = U^{\alpha \beta}_{\hat{\bf k}\hat{\bf q}} E^\beta(\hat{\bf q}) 
\end{align}
with the transformation matrix
\begin{align} \label{TrafoMatrix}
U_{\hat{\bf k}\hat{\bf q}} = 
\left(
\begin{array}{cc}
\cos 2\phi & \sin 2\phi \\
-\sin 2\phi & \cos 2\phi
\end{array}
\right),
\end{align}
where $\hat{\bf q}\hat{\bf k} = \cos \phi$ and $\hat{\bf p}\hat{\bf k} = \sin \phi$. $U$ describes a rotation but with angle $2\phi$ reflecting the invariance of the quadrupolar field under $\phi = \pi$ rotations.

\subsection{Quadrupolar polarization of free fermions}
\label{sec:Polarization}

\begin{figure}
\includegraphics[width=0.48\textwidth]{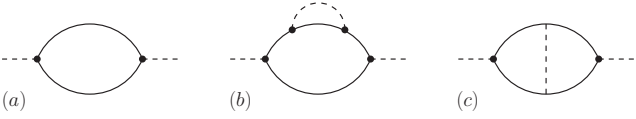} 
\caption{Contributions to the quadrupolar polarization of the Fermi liquid. Solid lines are fermion propagators, the dashed lines are susceptibilities of shear fluctuations, and the vertex is given by the quadrupolar gradient tensor $\hat{Q}_{ij}$. (a) polarization of free fermions, (b) and (c) lowest-order corrections.}
\label{fig:diagrams}
\end{figure}

The quadrupolar polarization of the fermions, see Fig.~\ref{fig:diagrams}(a), gives rise to a dynamical component of the shear susceptibility tensor $\chi_{ijkl} = \langle \sigma_{ij} \sigma_{kl} \rangle$,
\begin{align}
\chi^{-1}_{0 ijkl}({\bf q},i \Omega_n) =  -\frac{1}{F_2}
\mathds{1}_{ijkl} - \Pi^0_{ijkl}({\bf q},i\Omega_n),
\end{align}
where $\mathds{1}_{ijkl} = \frac{1}{2}\left(\delta_{ik} \delta_{jl}+\delta_{il}\delta_{jk} - \delta_{ij} \delta_{kl}\right)$ is the projection onto the $2d$ quadrupolar subspace and $\Pi^0$ is the lowest-order quadrupolar polarization of fermions,
\begin{align}
\Pi^0_{ijkl}({\bf q},i \Omega_n) =& -\frac{1}{\beta} \sum_{{\bf k}, \omega_n} {Q}_{ij}({\bf k}) {Q}_{kl}({\bf k})
\\\nn&
\times g_0({\bf k}+\frac{{\bf q}}{2},i \omega_n+i \Omega_n) g_0({\bf k}-\frac{{\bf q}}{2},i \omega_n).
\end{align}
In the basis (\ref{Eigenvectors}) for the shear modes, the susceptibility $\chi^{-1}_{\alpha\beta} = E^\alpha_{ij} \chi^{-1}_{ijkl} E^\beta_{kl}$ reads
\begin{align} \label{ShearSuscep}
\chi^{-1}_{\alpha\beta}({\bf q},i\Omega_n) =  - \frac{1}{F_2} \delta_{\alpha\beta} - \Pi^0_{\alpha \beta}({\bf q},i\Omega_n),
\end{align}
where $\alpha, \beta \in \{\parallel,\perp\}$. The polarization at zero external frequency and momentum is given by the density of states $\Pi^0_{\alpha \beta}(0,0) = \nu \delta_{\alpha \beta}$. The criterion $\chi^{-1}_{\alpha \beta}(0,0) =0$ for the Pomeranchuk instability to occur translates to\cite{Pomeranchuk54}
\begin{align}
\nu F_2 = -1.
\end{align}
The dynamic part of the polarization is also diagonal in the basis (\ref{Eigenvectors}). Its longitudinal part reduces to 
\begin{align} \label{Pol0LongDyn}
\left. \Pi^0_{\parallel\parallel}({\bf q},i\Omega_n)\right|_{\rm dyn} &= - i \Omega_n \nu \int_0^{2\pi} \frac{d \varphi}{2\pi} \frac{2\cos^2 2\varphi}{i \Omega_n - v_F q \cos \varphi} 
\nn\\& 
\approx - 2\nu \frac{|\Omega_n|}{v_F q}.
\end{align}
For the transversal part one obtains instead
\begin{align} \label{Pol0TransDyn}
\left. \Pi^0_{\perp\perp}({\bf q},i\Omega_n)\right|_{\rm dyn} &=- i \Omega_n \nu \int_0^{2\pi} \frac{d \varphi}{2\pi}
\frac{2\sin^2 2\varphi}{i \Omega_n - v_F q \cos \varphi}
\nn\\& 
 \approx - 4 \nu \left(\frac{\Omega_n}{v_F q}\right)^2.
\end{align}
The integral over the angle $\varphi$ takes into account the possible relative orientations of the mean Fermi velocity of the particle-hole pair and the momentum of the shear fluctuations, ${\bf q}$; the $\varphi$ dependence of the weight distinguishes the longitudinal from the transversal mode. The approximations in Eqs.~(\ref{Pol0LongDyn}) and (\ref{Pol0TransDyn}) are valid to lowest order in $\Omega_n / (v_F q)$.

\begin{figure}
\includegraphics[width=0.3\textwidth]{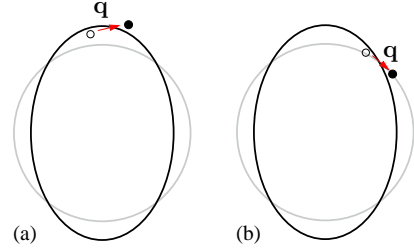} 
\caption{
Visualization of a quadrupolar fluctuation (black solid line) of the $2d$ isotropic Fermi sphere (gray solid line). (a) The longitudinal mode is damped by exciting particle-hole pairs close to the Fermi surface, Eq.~(\ref{Pol0LongDyn}). (b) For the transversal mode the momentum $\bf q$ of the fluctuation is close to a node such that there is not sufficient phase space for Landau damping, Eq.~(\ref{Pol0TransDyn}).
}
\label{fig:Damping}
\end{figure}

The origin of the different dynamics of the two modes is illustrated in Fig.~\ref{fig:Damping}. The quadrupolar fluctuation of the Fermi sphere cuts the undistorted sphere in four nodes. The nature of the Pomeranchuk dynamics depends on the relation between the location of these nodes and the direction of bosonic momentum {\bf q}. 
In the low-energy limit $\omega \ll v_F q$, the bosonic momentum $\bf q$ is approximately tangential to the Fermi surface. Near the antinodes, see Fig.~\ref{fig:Damping}(a), Landau damping is active and the longitudinal quadrupolar mode is damped by exciting particle-hole pairs, Eq.~(\ref{Pol0LongDyn}). Near the nodes, on the other hand, the lack of phase space for exciting particle-hole pairs results in an undamped dynamics for the transversal mode, Eq.~(\ref{Pol0TransDyn}), see Fig.~\ref{fig:Damping}(b).

In the lowest order in frequency and momentum, the shear susceptibility [Eq.~(\ref{ShearSuscep})] is diagonal in the basis (\ref{Eigenvectors}), $\chi^{-1}_{0\alpha\beta} = \delta_{\alpha\beta }\chi^{-1}_{0\alpha\alpha}$, with the two components,
\begin{subequations}
\label{ShearSuscep2}
\begin{align} 
\chi^{-1}_{0\parallel\parallel}({\bf q},i\Omega_n) &= r_0 + K q^2 + \eta \frac{|\Omega_n|}{v_F q},
\\
\chi^{-1}_{0\perp\perp}({\bf q},i\Omega_n) &= r_0 + K q^2 + \eta' \frac{\Omega_n^2}{(v_F q)^2},
\end{align}
\end{subequations}
where $r_0 = -1/(F_2) - \nu$, and $\eta = \eta'/2 = 2\nu$. Note that the Franck constant $K$, see Eq.~(\ref{EnergyFunctional}), is not determined by the low-energy properties of the model [Eq.~(\ref{model})] but depends on the details of the electron band structure. 

As advertised, from Eqs.~(\ref{ShearSuscep2}) it follows that the dynamics of the shear fluctuations are characterized by two different dynamical exponents that define the dispersion of the two eigenmodes, $\omega \sim q^z$, at criticality $r_0 \approx 0$. Whereas the damped longitudinal mode has $z_{>}=3$, the dynamical exponent of the undamped transversal mode is instead $z_<=2$. Note that symmetry ensures that both modes have the same tuning parameter $r_0$ even after including interaction corrections and they become simultaneously critical. The Pomeranchuk quantum critical point is thus a natural candidate to study the interplay of critical modes with different dynamics. 
 
\subsection{Corrections to the quadrupolar polarization of free fermions}

One may wonder whether the dynamics of the shear susceptibility [Eq.~(\ref{ShearSuscep2})] is robust or gets modified in higher order loop corrections. The damped longitudinal shear fluctuations couple back to the fermions resulting in singular self-energy corrections.\cite{Oganesyan01} The resulting non-Fermi liquid might screen the shear fluctuations in a different manner than the Fermi liquid maybe leading to a modification of the effective shear fluctuation dynamics. It was speculated in Ref.~\onlinecite{Woelfle07} that this feedback might alter the dynamics of the transversal mode and change its dynamical exponent.

A detailed analysis of the next-to-leading loop diagrams in Figs.~\ref{fig:diagrams}(b) and \ref{fig:diagrams}(c) is presented in Appendix \ref{app:FBModel}. We indeed find that by considering only the quadrupolar polarization diagram (b) containing the singular self-energy correction one would erroneously conclude that the dynamics of the transversal fluctuations gets modified from $z_< = 2$ to $z_< = 12/5$, see Eq.~(\ref{Result(b)}). However, it turns out that the contribution from the singular self-energy in diagram (b) is just canceled by the vertex correction displayed in Fig.~\ref{fig:diagrams}(c). 
Similar cancelations of singular self-energy and vertex contributions in the electron polarization are well known, for example, from the problem of electrons coupled to a Landau-damped gauge field.\cite{Lee89,Altshuler94,Kim94} The root of this cancellation is the Ward identity deriving from particle number conservation.\cite{Metzner98} 
Upon including both diagrams (b) and (c), we finally arrive at the result that the dynamics of the shear fluctuations [Eq.~(\ref{ShearSuscep2})] is robust with respect to the lowest-order corrections to the free fermion quadrupolar polarization. 
In the present problem of the Pomeranchuk instability, self-energy and vertex corrections thus have to be treated on the same footing in order to maintain the dynamics of the shear modes. Note that for this reason a simple version of Eliashberg theory, which neglects the vertex diagram Fig.~\ref{fig:diagrams}(c), is not applicable.

In the following, we will take the form of the dynamics for the shear fluctuations for granted that derive from the free-fermion polarization, Eqs.~(\ref{ShearSuscep2}). Using this dynamics, we then consider the effective Ginzburg-Landau theory for the shear modes in the spirit of Hertz.\cite{Hertz76}

\section{Ginzburg-Landau theory for Pomeranchuk quantum criticality}
\label{sec:BosonicTheory}

From the static energy functional [Eq.~(\ref{EnergyFunctional})] and the shear susceptibility [Eq.~(\ref{ShearSuscep2})] we can construct an effective Ginzburg-Landau theory for the shear modes, $\sigma_{ij}$, of the Fermi sphere,
\begin{align} \label{EffShearTheory}
\mathcal{S}[\sigma] = 
\int d\tau d^d x \left[
\frac{1}{2} \sigma_{ij} \chi_{ijkl}^{-1} \sigma_{kl} + \frac{u}{12} {\rm tr }\{ \sigma^4 \} \right].
\end{align}
We limit ourself to spatial dimensions $d=2$. As already mentioned in the context of Eq.~(\ref{EnergyFunctional}), a third-order term tr$\{\sigma^3\}$ vanishes  in $d=2$. Expanding the shear modes in the basis (\ref{Eigenvectors}) with respect to a fixed momentum direction, say $E^\alpha(\hat{e}_y)$ with det$(\hat{e}_y,\hat{e}_x) = 1$, the theory takes the form of a $\Phi^4$ theory of a two-component field $\Phi = (\phi_1, \phi_2)$,
\begin{align} \label{Phi^4-Theory}
\mathcal{S}[\Phi] = 
\int d\tau d^d x \left[
\frac{1}{2} \Phi^T \chi^{-1} \Phi + \frac{u}{4!} (\Phi^T \Phi)^2 \right].
\end{align}
For the susceptibility $\chi$, we use the free fermion form [Eq.~(\ref{ShearSuscep2})] derived in Sec.~\ref{sec:Polarization}. It is a $2\times2$ matrix given by 
\begin{align}\label{SuscTensor}
\chi^{-1}_{{\bf q},i\Omega_n} = 
U_{\hat{e}_y \hat{\bf q}}\, {\rm diag}\{\chi^{-1}_{\parallel\parallel{\bf q},i\Omega_n},\chi^{-1}_{\perp\perp{\bf q},i\Omega_n}\}  U^T_{\hat{e}_y \hat{\bf q}},
\end{align}
where the transformation matrix $U_{\hat{e}_y \hat{\bf q}}$ is defined in Eq.~(\ref{TrafoMatrix}). The longitudinal and transversal components of the susceptibility are 
%
\begin{align} \label{LongSuscep}
\chi^{-1}_{\parallel\parallel}({\bf q},i\Omega_n) &= r + q^2 + \eta_> \frac{|\Omega_n|}{q},
\\\label{TransSuscep}
\chi^{-1}_{\perp\perp}({\bf q},i\Omega_n) &= r + q^2 + \eta^2_< \frac{\Omega_n^2}{q^2}
\end{align}
with a conveniently rescaled momentum and temperature; 
$\eta_>$ and $\eta_<$ are dimensionless (positive) parameters. Note that we could have chosen units such that either $\eta_>$ or $\eta_<$ is unity so that one of them is in fact {\em redundant}; we deliberately keep here both explicitly for later convenience. After rescaling, the dimensionless length and temperature are measured in units of $[\sqrt{K/\nu}]$ and $[v_F \sqrt{\nu/K}]$, respectively, where $K$ is the Franck constant, $\nu$ is the fermionic density of states, and $v_F$ is the Fermi velocity. \hspace*{-.3em}The parameter $r$ controls the distance to the Pomeranchuk \hspace*{.2em}transition \hspace*{.2em}and \hspace*{.2em}the interaction amplitude is $u$.

\subsection{Perturbative renormalization group at $T=0$}
\label{sec:RG}

\begin{figure}
\includegraphics[width=0.3\textwidth]{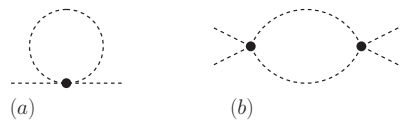} 
\caption{One-loop corrections for the effective theory [Eq.~(\ref{EffShearTheory})] of the shear modes. 
}
\label{fig:Bosons}
\end{figure}

The transversal component of the shear modes with susceptibility $\chi_{\perp\perp}$ has dynamical exponent $z_< = 2$ and thus an effective dimension $d+z_<$ equal to the upper critical dimension $d^+_c = 4$ of the $\Phi^4$ theory [Eq.~(\ref{Phi^4-Theory})]. Perturbation theory in the interaction $u$ at temperature $T=0$ is therefore accompanied by logarithmic singularities which we sum up with the help of an one-loop RG treatment. 

Consider the perturbative corrections in Fig.~\ref{fig:Bosons}. For their evaluations the angular averages of shear mode propagators given in the Appendix~\ref{sec:Averages} are needed. The one-loop self-energy diagram (a) leads to a shift of the tuning parameter $r \to r + \delta r$,
\begin{align} \label{MassCorr}
\delta r = 
\frac{u}{3} \frac{1}{\beta} \sum_{{\bf q},\Omega_n} 
\left( \chi_{\parallel\parallel {\bf q}, i\Omega_n} +  
\chi_{\perp\perp {\bf q}, i\Omega_n}\right).
\end{align}
Similarly, the vertex correction (b), $u \to u + \delta u$, is given by 
\begin{align} \label{VerCorr}
\delta u = -\frac{2 u^2}{4!} \frac{1}{\beta}\sum_{{\bf q},\Omega_n} \left[
9 \left( \chi^2_{\parallel\parallel {\bf q}, i\Omega_n} +  
\chi^2_{\perp\perp {\bf q}, i\Omega_n} \right) 
\right.
\\\nn
\left. +
2 \chi_{\parallel\parallel {\bf q}, i\Omega_n} \chi_{\perp\perp {\bf q}, i\Omega_n} \right].
\end{align}
As usual, the renormalization of the control parameter [Eq.~(\ref{MassCorr})] yields a nonuniversal shift, which depends on the UV cutoff of the integrals. In the following, we assume that such finite renormalizations have been already absorbed by appropriate counterterms.\cite{ZinnJustin} In addition, the diagram (a) gives a universal albeit logarithmically divergent correction proportional to the control parameter, $r$, itself arising from the transversal $z_< = 2$ quadrupolar mode,
%
\begin{align} \label{MassLog}
\frac{\partial \delta r}{\partial r} \approx - \frac{u}{3} \frac{1}{\beta} \sum_{{\bf q},\Omega_n} 
\chi^2_{\perp\perp {\bf q}, i\Omega_n} 
\approx - \frac{u}{24 \pi \eta_< } \log\left(\frac{\Lambda}{\sqrt{r}} \right).
\end{align}
In \hspace*{-.2em}the \hspace*{-.2em}last \hspace*{-.2em}expression, \hspace*{-.2em}we \hspace*{-.2em}evaluated \hspace*{-.2em}the \hspace*{-.2em}integrals \hspace*{-.2em}at \hspace*{-.2em}$T\!=\!0$ with a sharp momentum cutoff $\Lambda$ and kept only the leading logarithmic term. Similarly, we get for the vertex correction
%
\begin{align}\label{ULog}
\delta u \approx 
-\frac{2 u^2}{4!} 9
\frac{1}{\beta} \sum_{{\bf q},\Omega_n} 
\chi^2_{\perp\perp {\bf q}, i\Omega_n} 
\approx
- \frac{3 u^2}{32\pi \eta_<} \log \left(\frac{\Lambda}{\sqrt{r}} \right).
\end{align}

Equations.~(\ref{MassLog}) and (\ref{ULog}) are the only logarithmic singularities encountered in lowest-order perturbation theory. In order to treat these singularities, we apply the renormalization group and perturbatively integrate out modes within a momentum shell $(\Lambda/b,\Lambda)$ with $\log b \ll 1$. After such a single RG step, the momentum cutoff of the perturbatively corrected theory is restored to $\Lambda$ by rescaling momenta $q \to q/b$ and frequencies 
$\omega \to \omega/b^{z}$ with some dynamical exponent $z$. As we treat a theory with multiple dynamical scales, one may wonder what kind of value to choose for $z$ in this RG process. It is instructive to leave the value for $z$ here unspecified in order to discuss later possible choices. Of course, the result for physical observables will turn out to be independent of the precise choice for $z$. Using the renormalization-group conditions that the prefactor in front of the momentum dependence, $q^2$, of the susceptibilities, Eqs.~(\ref{LongSuscep}) and (\ref{TransSuscep}), remains unity, the perturbative renormalized parameters can then be read off. The scaling of the dynamical parameters, $\eta_<$ and $\eta_>$, at one-loop order is simply obtained from power counting
\begin{subequations}
\label{etaEqs}
\begin{align} \label{eta>RG}
\frac{\partial \eta_>}{\partial \log b} = (z_> - z) \eta_>,
\\ \label{eta<RG}
\frac{\partial \eta_<}{\partial \log b} = (z_< - z) \eta_<,
\end{align}
\end{subequations}
where again $z_< = 2$ and $z_> = 3$. The one-loop corrections to tree-level scaling for the control parameter, $r$, and the interaction, $u$, at $T=0$ follow directly from Eqs.~(\ref{MassLog}) and (\ref{ULog}),
\begin{subequations}
\label{RGEqs}
\begin{align} 
\label{RGEqMass}
\frac{\partial r}{\partial \log b} &= \left(2 - \frac{1}{24\pi \eta_<} u \right) r, 
\\ 
\label{RGQuarticCoup}
\frac{\partial u}{\partial \log b} &= (4 - d - z) u - \frac{3}{32\pi \eta_<}  u^2,
\end{align}
\end{subequations}
with $d=2$. These RG equations are characteristic for the universality class of the Pomeranchuk transition 
and are an important result of this work. 

Let us remark on the ambiguity with respect to the choice of $z$, that determines the tree-level scaling in Eqs.~(\ref{etaEqs}) and (\ref{RGEqs}). 
First of all, note that for all choices of $z>2$ the tree-level scaling suggests that the quartic coupling $u$ as well as the dynamical parameter $\eta_<$ are irrelevant.\cite{Oganesyan01} Nevertheless, we know from the perturbative results, Eqs.~(\ref{MassLog}) and (\ref{ULog}), that there exists a nontrivial marginal operator. Indeed, closer inspection of the RG equations readily shows that the quartic coupling enters the RG flow only in the combination 
\begin{align}
\tilde{u} \equiv \frac{u}{\eta_<}.
\end{align}
Correspondingly, the Eqs.~(\ref{RGEqs}) can be combined with Eq.~(\ref{eta<RG}) to give
\begin{subequations}
\label{RGEqs2}
\begin{align} 
\label{RGEqMass2}
\frac{\partial r}{\partial \log b} &= \left(2 - \frac{1}{24\pi} \tilde{u} \right) r, 
\\ 
\label{RGQuarticCoup2}
\frac{\partial \tilde{u}}{\partial \log b} &= - \frac{3}{32\pi}  \tilde{u}^2.
\end{align}
\end{subequations}
The seemingly irrelevant parameters $u$ and $\eta_<$ thus conspire to yield in fact a marginal operator with amplitude $\tilde{u} = u/\eta_<$. 
Note, in particular, that Eqs.~(\ref{RGEqs2}) are {\em independent} of the specific choice of $z$. 

The freedom to choose the exponent $z$ can now be exploited to simplify the RG flow in Eqs.~(\ref{etaEqs}). There are two evident choices, $z=z_<$ or $z=z_>$, that eliminates the flow of one redundant scaling variable, i.e., either $\eta_<$ or $\eta_>$, respectively. For $z=z_<$ the remaining flow of the dynamical parameter $\eta_>$ is unbounded, $\eta_> \to \infty$ whereas for $z=z_>$ the other parameter flows to zero, $\eta_< \to 0$. That means that depending on the choice of $z$ either $1/\eta_>$ or $\eta_<$ is irrelevant in the RG sense. 
Taken together, the RG flow has either the critical fixed point $r=\tilde{u}=1/\eta_>=0$ or $r=\tilde{u}=\eta_<=0$, respectively. However, note that both fixed points are only of limited use when computing physical properties. The reason is that both, $1/\eta_>$ and $\eta_<$, are in fact {\em dangerously} irrelevant for the respective choices of $z$. When considering, e.g., the critical thermodynamics, it turns out that neither $1/\eta_> \to 0$ nor $\eta_< \to 0$ is a well-defined limit, which is the very definition of a dangerously irrelevant variable,\cite{Cardy} see Appendix~\ref{app:Limits}. Nevertheless, let us stress that any choice of $z$ for the RG yields, of course, the same result when computing critical properties of physical observables.

The Eqs.~(\ref{RGEqs2}) are readily solved. The running coupling $\tilde{u}(b)$ is given by
\begin{align} \label{RunningVertex}
\tilde{u}(b) = \frac{32 \pi}{3} \frac{1}{\log( \bar{\Lambda} b/\Lambda)}
\end{align}
with $\bar{\Lambda} = \Lambda e^{32 \pi \eta_</ (3 u)}$. The scale dependence of the coupling $\tilde{u} = u/\eta_<$ can be interpreted as an effective quartic vertex, 
\begin{align} \label{EffVertex}
\lefteqn{u_{\rm eff}(q,\omega,T,\xi) =}
\\\nn&\qquad \frac{32 \pi \eta_<}{3} \frac{1}{\log (\bar{\Lambda}/{\rm max}\{q , \omega^{1/z_<},T^{1/z_<}, 1/\xi\})},
\end{align}
that depends on momentum, frequency, temperature and correlation length $\xi$.
The full quartic vertex generally depends on four momenta and frequencies. However, to logarithmic \hspace*{.3em}accuracy \hspace*{.3em}we \hspace*{.3em}can \hspace*{.3em}take \hspace*{.3em}$q = {\rm max}\{q_i\}$ \hspace*{.3em}and $\omega = {\rm max}\{\omega_i\}$. It is important to note that it is the dynamical exponent of the transversal mode $z_< = 2$ according to which the momentum scale is translated to energies in Eq.~(\ref{EffVertex}) as it is this very same mode that induces the logarithmic dependence of the vertex. 
 
\subsection{Correlation length}

We compute the dependence of the correlation length $\xi$ on the control parameter, $r$, and temperature, $T$. We distinguish between the regimes (I), (II), and (III) of the phase diagram in Fig.~\ref{fig:PD}. Whereas in regimes (I) and (II) the correlation length is determined by the zero-temperature flow of the control parameter, Eq.~(\ref{RGEqMass}), it is dominated by temperature, $T$,  in regime (III). Remarkably, this latter $T$ dependence is found to be universal independent of the quartic interaction $u$. This universality arises from a peculiar interplay of logarithmic singularities of different origin.

\subsubsection{Correlation length in regime (I)}

Consider first the correlation length at temperature $T=0$, where it is determined by the RG flow of the control parameter [Eq.~(\ref{RGEqMass})]. With the help of expression (\ref{RunningVertex}) for the running quartic coupling, we can solve for the running $r(b)$ from which follows the correlation length at zero temperature $\xi^{-2}|_{T=0} = r(b)/b^2|_{b = \Lambda \xi}$,
\begin{align} \label{Gap}
\left.\xi^{-2}\right|_{T=0} \sim \frac{r}{\left(\log\frac{\bar{\Lambda}}{\sqrt{r}} \right)^{4/9}}.
\end{align} 
This \hspace*{-.1em}logarithmic \hspace*{-.1em}enhancement \hspace*{-.1em}of \hspace*{-.1em}the \hspace*{-.1em}correlation \hspace*{-.1em}length \hspace*{-.1em}was already \hspace*{-.1em}advertised \hspace*{-.1em}in \hspace*{-.1em}Eq.~(\ref{Gap-1}). \hspace*{-.1em}The \hspace*{-.1em}exponent \hspace*{-.1em}$4/9$ \hspace*{-.1em}is \hspace*{-.1em}characteristic \hspace*{-.1em}for \hspace*{-.1em}the \hspace*{-.1em}Pomeranchuk \hspace*{-.1em}universality \hspace*{-.1em}class. \hspace*{-.1em}It \hspace*{-.1em}is \hspace*{-.1em}instructive \hspace*{-.1em}to \hspace*{-.1em}compare \hspace*{-.1em}this \hspace*{-.1em}exponent \hspace*{-.1em}with \hspace*{-.1em}the \hspace*{-.1em}value \hspace*{-.1em}obtained \hspace*{-.1em}from \hspace*{-.1em}a \hspace*{-.1em}$O(N)$-symmetric $\Phi^4$ theory \hspace*{-.1em}with \hspace*{-.1em}a \hspace*{-.1em}propagator \hspace*{-.1em}containing \hspace*{-.1em}only \hspace*{-.1em}the $z_< = 2$ dynamics, $\chi^{-1}_{\alpha \beta} = \delta_{\alpha \beta} \chi^{-1}_{\perp\perp}$ and $\alpha,\beta = 1,\dots, N$. In such a case, one would obtain the exponent $(N+2)/(N+8)$ instead of $4/9$. In particular, note that the Ising ($N=1$) value $3/9$ as well as the XY ($N=2$) exponent $4/10$ differ from the value we obtain for the Pomeranchuk transition. The difference to the Ising exponent $3/9$ clearly indicates that the model [Eq.~(\ref{Phi^4-Theory})] cannot effectively be reduced to a $\Phi^4$ theory of a single-component field with dynamics $z_<=2$ as one might naively expect. The origin of the anomalous value $4/9$  of the exponent in Eq.~(\ref{Gap}) can be traced back to the result of angular averages over shear-mode propagators, Eqs.~(\ref{AngAverProd}), and thus reflects the topology of the shear field.

At finite temperatures, the correlation length also acquires a $T$ dependence. In the low-temperature regime (I) of the phase diagram in Fig.~\ref{fig:PD}, this $T$ dependence is only a sub-leading correction to Eq.~(\ref{Gap}) and can be computed within renormalized perturbation theory from the diagram in Fig.~\ref{fig:Bosons}(a). It is the thermal occupation of $z_>=3$ fluctuations that yields the dominant $T$ correction, which is of Fermi-liquid type, $\xi^{-2} - \xi^{-2}_{T=0} \sim \mathcal{O}(u_{\rm eff} T^2 \xi^3)$, where $u_{\rm eff} \sim \tilde{u}(\Lambda \xi)$ is the quartic coupling [Eq.~(\ref{RunningVertex})] at the scale $b=\Lambda \xi$.

\subsubsection{Correlation length in regimes (II) and (III)}
\label{sec:CorrFiniteT}
 
The temperature dependence of the correlation length in regimes (II) and (III) arises from a subtle interplay of logarithmic IR singularities driven by quantum and classical fluctuations. In order to illustrate this peculiar low-energy structure of the theory, consider first the $T$ dependence of the control parameter deriving from the perturbative one-loop correction (a) in Fig.~\ref{fig:Bosons}, 
\begin{align} \label{HartreeTneq0}
\delta r - \left. \delta r \right|_{T=0}  
&= \frac{u}{3} \int_0^\infty \frac{dq\,q}{2\pi}\, 
\int_{0}^\infty \frac{d\omega}{\pi} \left( \coth \frac{\omega}{2T} -1\right)
\nn\\
&\quad \times
{\rm Im} \left\{\chi^{\parallel\parallel}_{q, \omega+i 0} +  
\chi^{\perp\perp}_{q, \omega+i0} \right\}.
\end{align}
As before, it turns out that it is the contribution of the longitudinal susceptibility $\chi^{\parallel\parallel}$ that gives the leading $T$ dependence in regimes (II) and (III). Furthermore, in $d=2$ the integral [Eq.~(\ref{HartreeTneq0})] is dominated by the limit of small frequencies $\omega$ in these regimes, i.e., the classical limit of the hyperbolic $\coth$ function corresponding to the contribution of the Matsubara zero mode. Generally, the classical limit is applicable whenever the frequency is smaller than temperature, $\omega < T$. Using the scaling relation, $\omega \sim q^{z_>}$ this translates to a regime of momenta, $q < {\xi^>_T}^{-1}$, with the thermal length $\xi^>_T \sim T^{-1/z_>}$, where the associated fluctuations have a classical character, see Fig.~\ref{fig:ModeMomenta},
\begin{align}\label{HartreeTneq0-2}
\delta r - \left. \delta r \right|_{T=0}  \approx \frac{u T}{3} 
\int_{0}^{{\xi_T^>}^{-1}} \frac{dq\,q}{2\pi}\, \chi^{\parallel\parallel}(q, \omega=0).
\end{align}

In the classical regime, the renormalized susceptibility reduces to $\chi^{-1}_{\parallel\parallel}(q) = \xi^{-2}+q^2$, and the momentum integral in Eq.~(\ref{HartreeTneq0-2}) will be cut off by the inverse correlation length $1/\xi$ at small $q$. Anticipating that the correlation length itself is either given by $\xi \sim 1/\sqrt{T}$ in regime (III) or approximately by Eq.~(\ref{Gap}) in regime (II), the momentum integral in Eq.~(\ref{HartreeTneq0-2}) extends over a parametrically large range $1/\xi < q < T^{1/3}$, where it is logarithmically enhanced. However, in this momentum range the transversal mode still possesses its quantum character, see Fig.~\ref{fig:ModeMomenta}. As we showed in the previous section, for momenta larger than the inverse correlation length, $q > 1/\xi$, the vertex acquires an effective scale dependence due to the repeated scattering of quantum fluctuations in the transversal channel, see Eq.~(\ref{EffVertex}). 
As the integral in Eq.~(\ref{HartreeTneq0-2}) is logarithmically enhanced, the temperature correction turns out to be 
sensitive to this scale dependence of the vertex which becomes apparent in higher-order loop corrections. The scale dependent vertex [Eq.~(\ref{EffVertex})] is thus probed by the classical fluctuations in the longitudinal channel with momenta belonging to the extended quantum-to-classical crossover regime $(T^{1/2}, T^{1/3})$ of Fig.~\ref{fig:ModeMomenta}.
In order to determine the correlation length it is therefore insufficient to limit oneself to the lowest-order diagram (a). 

Instead, we apply the RG generalized to finite $T$, see Appendix \ref{app:RGT>0}. When the cutoff is lowered down to a value on the order of the inverse thermal length ${\xi^>_T}^{-1}$ corresponding to a scale $b \sim \Lambda \xi^>_T$, the contribution arising from the integral [Eq.~(\ref{HartreeTneq0-2})] translates to a modification of the zero-temperature flow in Eq.~(\ref{RGEqMass}). The classical longitudinal fluctuations give rise to an accelerated flow of the running control parameter $r(b)$ setting in at the scale $b \sim \Lambda \xi^>_T$,
%
\begin{align} \label{accRGFlow}
\frac{\partial r}{\partial \log b} = \left(2 - \frac{1}{24 \pi \eta_<} u \right) r + \frac{1}{6\pi} u T\, \Theta_{b - \Lambda \xi^>_T}
\end{align}
where $\Theta_x = 1$ if $x>0$ and $0$ otherwise. The flow of the quartic coupling is still given by Eq.~(\ref{RGQuarticCoup}) and the temperature has the scale dependence $T(b) = T b^{z}$. Solving for $r(b)$ and identifying the correlation length with $\xi^{-2} = r(b)/b^2|_{b = \Lambda \xi}$, we get the following implicit equation for $\xi$ valid in regimes (II) and (III) of the phase diagram in Fig.~\ref{fig:PD},
%
\begin{align} \label{CorrelationLength} 
\xi^{-2} &= r  \left(\frac{\log \bar{\Lambda}/\Lambda}{\log \bar{\Lambda} \xi}\right)^{4/9}
+ 4 \eta_< T \left(1 -  \left(\frac{\log \bar{\Lambda}\xi^{>}_T}{\log \bar{\Lambda} \xi}\right)^{4/9}   \right)
\end{align}
with $\xi^{>}_T \sim T^{-1/3}$. The first term dominates in regime (II) and yields a value for $\xi$ that coincides with expression (\ref{Gap}). The second term originates from the accelerated RG flow due to the classical fluctuations in the longitudinal channel. This acceleration acts on scales in the window $\Lambda \xi^>_T < b < \Lambda \xi$, which can be identified with the momentum range in Fig.~(\ref{fig:ModeMomenta}) where the quantum regime of the transversal mode overlaps with the classical regime of the longitudinal fluctuations. This $T$ dependence dominates the correlation length in the quantum critical regime (III).

At criticality $r=0$, Eq.~(\ref{CorrelationLength}) is easily solved iteratively and we get a universal asymptotic behavior at low temperatures,
%
\begin{align} \label{UniversalXi}
\left.\xi^{-2}\right|_{r=0} = \mathcal{C}\,  \eta_< T,\qquad \mathcal{C} = 4 \left[1 - \left(\frac{2}{3}\right)^{4/9}\right].
\end{align}
This reduces to Eq.~(\ref{CorrLengthUniversalT}) in Sec.~\ref{sec:SummaryResults} after exploiting the freedom to choose temperature units such that 
$\eta_< = 1$, cf.~the remark below Eq.~(\ref{TransSuscep}). Note that the temperature dependence of $\xi$ is here independent of the bare value of the quartic coupling $u$. It enters the implicit Eq.~(\ref{CorrelationLength}) via $\bar{\Lambda}$ defined below Eq.~(\ref{RunningVertex}) but drops out in the asymptotic limit $T\to 0$. This universal temperature dependence  originates from an intricate interplay of quantum fluctuations in the transversal channel and the classical shear fluctuations in the longitudinal channel.

\subsubsection{Ginzburg temperature}

Our perturbative treatment finally breaks down when classical loop corrections start to dominate. This is the case when the Ginzburg parameter is on the order of one, $\mathcal{G} = u_{\rm eff} T / \xi^{-2}$, where $u_{\rm eff} \sim \tilde{u}(\Lambda\xi)$ is the running vertex [Eq.~(\ref{RunningVertex})] at the scale $b=\Lambda \xi$. The criterion $\mathcal{G} \sim 1$ identifies the Ginzburg temperature $T_G(r)$. In the asymptotic low-temperature limit, the Ginzburg temperature is given by
\begin{align}
T_G(r) \sim - \frac{r}{\left(\log \frac{\bar{\Lambda}}{\sqrt{|r|}} \right)^{4/9}}
\end{align}
and is shown in Fig.~\ref{fig:PD}.

\subsection{Critical thermodynamics}
\label{sec:CrThermodynamics}

The presence of two dynamical scales gives rise to two crossover lines in the phase diagram, see Fig.~\ref{fig:PD}. In the present section, we investigate how these crossovers affect the critical thermodynamics. 

The critical free energy, $\mathcal{F}_{\rm cr}$, depends on the control parameter, $r$, and temperature, $T$. We consider the thermodynamic quantities that 
correspond to the three second-order derivatives of $\mathcal{F}_{\rm cr}$ with respect to $r$ and $T$. The derivative, $\gamma_{\rm cr} = - \partial^2 \mathcal{F}_{\rm cr}/\partial T^2$ is the specific heat coefficient. We discuss the derivatives with respect to $r$ in the language of a pressure-tuned quantum phase transition. If the control parameter depends smoothly on pressure, $r = r(p)$, derivatives with respect to $r$ effectively correspond to derivatives with respect to $p$.\cite{Zhu03} Consequently, the quantity $\alpha_{\rm cr} = \partial^2 \mathcal{F}_{\rm cr}/(\partial T\partial r)$ can be identified with the thermal expansion and $\kappa_{\rm cr} = - \partial^2 \mathcal{F}_{\rm cr}/\partial r^2$ parallels the critical contribution to the compressibility. 

It is instructive to discuss the asymptotic behavior close to criticality within a renormalized Gaussian model, where the free energy is given by
\begin{align} \label{GaussEnergy}
\mathcal{F}_{\rm ren. Gauss} = \frac{1}{2} 
\int \frac{d^2 q}{(2\pi)^2} \frac{1}{\beta}\sum_{\Omega_n} 
\log {\rm det} \chi^{-1}(q,i\Omega_n)|_{r \to \xi^{-2}}
\end{align}
with the susceptibility tensor of Eq.~(\ref{SuscTensor}) but with the control parameter replaced by the correlation length, $r \to \xi^{-2}$. The approximate expression (\ref{GaussEnergy}) for the free energy accounts for most of the critical behavior except the thermal expansion in regimes (II) and (III), and the compressibility in regime (I) of Fig.~\ref{fig:PD}. In order to obtain the correct asymptotic behavior of the latter, we have to resort to a proper RG improved treatment. The quantum fluctuations of the $z_< = 2$ mode lead here to scale-dependent vertices that are beyond the renormalized Gaussian form of Eq.~(\ref{GaussEnergy}). Below, we only present the result of the asymptotic critical behavior and refer the reader to Appendix \ref{app:RGT>0} for its derivation.

\subsubsection{Specific heat coefficient}

The critical specific heat, $\gamma_{\rm cr} = - \partial^2 \mathcal{F}_{\rm cr}/\partial T^2$, is dominated by the damped $z_>=3$ shear fluctuations because their available phase space is larger. 
Their contribution to the specific heat can be derived within the renormalized Gaussian model with a free energy given by Eq.~(\ref{GaussEnergy}). In the low-temperature regime (I) of Fig.~\ref{fig:PD}, the critical specific heat is temperature independent in the limit $T \to 0$, 
\begin{align}
\gamma_{\rm cr} \sim \xi 
\sim \frac{\left(\log\frac{\bar{\Lambda}}{\sqrt{r}}\right)^{2/9}}{r^{1/2}}
\qquad {\rm (I)}.
\end{align}
In the overlap (II) and quantum critical regime (III), we get
\begin{align} \label{SpecHeat-III}
\gamma_{\rm cr} \sim T^{-1/3} \qquad {\rm (II)\quad  and\quad  (III)}.
\end{align}
The undamped $z_<=2$ mode adds a sub-leading correction, which is on the order $\mathcal{O}(T \xi^2)$ in regimes (I) and (II), and $\mathcal{O}(1)$ in regime (III). Only the subleading corrections to $\gamma_{\rm cr}$ distinguish the regimes (II) and (III). Note that the $z_< = 2$ mode, nevertheless, leaves its traces in $\gamma_{\rm cr}$ via the logarithmic renormalization entering the correlation length in regime (I).

\subsubsection{Thermal expansion}

The dependences of the thermal expansion, defined as $\alpha_{\rm cr} = \partial \mathcal{F}_{\rm cr}/(\partial r\partial T)$, is more delicate. It is again the large phase space of the $z_> = 3$ mode that is the main driving force. In the low-$T$ regime (I), $\alpha_{\rm cr}$ can still be understood within the renormalized Gaussian model [Eq.~(\ref{GaussEnergy})],
\begin{align}
\alpha_{\rm cr} \sim T \xi^3 \frac{\partial \xi^{-2}}{\partial r} \sim \frac{T}{r^{3/2}}
\left(\log\frac{\bar{\Lambda}}{\sqrt{r}}\right)^{2/9}
\qquad {\rm (I)}.
\end{align}
This leading contribution is attributed to the $z_>=3$ part of Eq.~(\ref{GaussEnergy}). The $z_< = 2$ mode again enters here indirectly in the form of the logarithmically renormalized correlation length. 

Interestingly, the critical thermal expansion, $\alpha_{\rm cr}$, with-in the regimes (II) and (III) is more involved. Let us first discuss the result one would obtain from the renormalized Gaussian model [Eq.~(\ref{GaussEnergy})]. In $d=2$, it is the classical Matsubara zero mode that gives the dominating contribution to $\alpha_{\rm cr}$,
\begin{align} \label{AlphaGaussMot}
\alpha_{\rm cr} \sim \int_0^{{\xi_T^>}^{-1}} dq \frac{q}{\xi^{-2} + q^2} 
\frac{\partial \xi^{-2}}{\partial r}.
\end{align}
We only retained the part due to the classical $z_> =3$ mode here; the additive classical $z_< = 2$ contribution is cut off at the smaller momentum scale ${\xi_T^<}^{-1}$ instead and turns out to be subleading. The derivative $\partial \xi^{-2}/\partial r$ can be interpreted as an effective vertex that is, in general, scale dependent due to the RG flow of the control parameter discussed in the context of Eq.~(\ref{RGEqMass}). As the momentum integral in Eq.~(\ref{AlphaGaussMot}) covers the overlap regime $(1/\xi, 1/\xi_T^>)$, where the $z_< = 2$ fluctuations are still quantum, cf.~Fig.~\ref{fig:ModeMomenta}, this scale dependence is fully developed in the important integration range and has to be taken into account.
Similarly as for the correlation length, the quantum fluctuations of the $z_<=2$ mode here give rise to a scale dependent vertex that is probed by the classical fluctuations of the $z_>=3$ mode. The critical thermal expansion is thus beyond renormalized Gaussian but can be obtained from Eq.~(\ref{AlphaGaussMot}) by replacing $\partial \xi^{-2}/\partial r \to 1/(\log(\bar{\Lambda}/q))^{4/9}$. So we finally get to logarithmic accuracy
\begin{align} \label{ThermExpQC}
\alpha_{\rm cr} \sim \int_{\xi^{-1}}^{{\xi_T^>}^{-1}} 
\frac{dq}{q} \frac{1}{[\log(\bar{\Lambda}/q)]^{4/9}}
\qquad {\rm (II)\;\;  and\;\;  (III)}
\end{align}
whose asymptotic limit depends on whether the correlation length is temperature dominated, (III), or $r$ dominated, (II). To summarize, we have
\begin{align}
\alpha_{\rm cr} \sim \left\{
\begin{array}{lc}
\frac{T}{r^{3/2}}
\left(\log\frac{\bar{\Lambda}}{\sqrt{r}}\right)^{2/9}& \quad {\rm (I)}\\ 
\left(\log\frac{\bar{\Lambda}}{\sqrt{r}}\right)^{5/9}& \quad {\rm (II)}\\
\left(\log\frac{\bar{\Lambda}}{\sqrt{T}}\right)^{5/9} & \quad {\rm (III)} 
\end{array}\right.
\end{align}
The thermal expansion thus changes its behavior at both crossovers in Fig.~\ref{fig:PD}.

\subsubsection{Compressibility}

The scale-dependent vertex in Eq.~(\ref{ThermExpQC}) is also important for the critical compressibility, $\kappa_{\rm cr} = - \partial^2 \mathcal{F}_{\rm cr}/\partial r^2$. 
In the low-temperature (I) and overlap regime (II), the Gaussian free energy [Eq.~(\ref{GaussEnergy})] predicts a logarithmic divergence of the critical compressibility due to the undamped $z_<=2$ shear fluctuations
\begin{align} \label{CompLowT}
\kappa_{\rm cr} \sim \int_{\xi^{-1}} \frac{dk}{k} \left(\frac{\partial \xi^{-2}}{\partial r}\right)^2.
\end{align}
The \hspace*{-.2em}correct \hspace*{-.2em}behavior \hspace*{-.2em}again \hspace*{-.2em}follows \hspace*{-.2em}upon \hspace*{-.2em}RG \hspace*{-.2em}improvement by \hspace*{-.2em}replacing \hspace*{-.2em}the \hspace*{-.2em}vertex $\partial \xi^{-2}/\partial r \!\to\! 1/[\log(\bar{\Lambda}/k)]^{4/9}$,
\begin{align} \label{Comp_I+II}
\kappa_{\rm cr} \sim \left(\log\frac{\bar{\Lambda}}{\sqrt{r}}\right)^{1/9}
\qquad {\rm (I)\quad  and\quad  (II)}.
\end{align}
Alternatively, this result can be derived directly from the non-trivial RG flow of the critical free energy as explained in Appendix \ref{app:RGT>0}. Note that the logarithmic divergence of the compressibility as $r\to 0^+$ indicates that a coupling of other degrees of freedom to the square of the shear field tr$\{\sigma^2\}$ might be nonperturbative.\cite{Anfuso08}

In the quantum critical regime (III), on the other hand, the correct asymptotics can again be obtained from Eq.~(\ref{GaussEnergy}),
\begin{align} \label{compressQCR}
\kappa_{\rm cr} 
\sim
\left(\log\frac{\bar{\Lambda}}{\sqrt{T}}\right)^{-8/9}
\qquad {\rm (III)}. 
\end{align}
Both modes contribute to this logarithmic dependence of the compressibility. While the $z_>=3$ fluctuations contribute only with their classical component, i.e., Matsubara zero mode, remarkably, all Matsubara frequencies of the $z_<=2$ fluctuations add to the logarithmic dependence [Eq.~(\ref{compressQCR})] because $\xi \sim \xi^<_T$ in regime (III). 

So we find that the asymptotic behavior of the critical compressibility is only sensitive to the (II)/(III) crossover.

\section{Summary}
\label{sec:discussion} 

We considered the problem of multiscale quantum criticality by studying 
the effective bosonic model of a quadrupolar Pomeranchuk instability in an isotropic metal in spatial dimension $d=2$. This theory consists of two critical bosonic shear modes of the Fermi sphere characterized by different dynamics, resulting in a quantum criticality with multiple scales. While the longitudinal shear fluctuations are Landau damped and have a critical dynamical exponent $z_>=3$, the transversal modes are undamped with $z_<=2$.\cite{Oganesyan01}

First, we studied the case of zero temperature. There we found that the low-energy properties
are governed by the mode with the smaller effective dimension $d+z_<$, i.e., the transversal shear fluctuations. Since that effective dimension is equal to the upper critical dimension, $d^+_c =4$, of the effective $\Phi^4$ theory, its self-interactions give rise to logarithmic singularities in perturbation theory and, as a consequence, to a nontrivial RG flow of the theory, Eqs.~(\ref{RGEqs}). The interaction amplitude is found to be marginally irrelevant in the RG sense so that the low-energy quantum critical fixed point is Gaussian. Nevertheless, as usual the slow flow of the interaction amplitude toward weak coupling gives rise to qualitative corrections to Gaussian behavior. These corrections consist of, e.g., a logarithmic enhancement of the correlation length, Eq.~(\ref{Gap}) and a logarithmically diverging compressibility, Eq.~(\ref{Comp_I+II}), whose functional form is characteristic for the Pomeranchuk universality class. 

Second, we investigated the critical properties at finite temperatures. A salient feature of quantum criticality, in general, is the quantum-to-classical crossover\cite{Sachdev, ZinnJustin,Millis93,Sachdev97} associated with the thermal length, $\xi_T \sim T^{-1/z}$. In the language of crossover theory,\cite{Nelson75} the renormalization-group flow of a quantum critical model on short scales is \hspace*{-.2em}first \hspace*{-.2em}governed \hspace*{-.2em}by \hspace*{-.2em}the \hspace*{-.2em}primary \hspace*{-.2em}quantum \hspace*{-.2em}critical, $T\!=\!0$, fixed point, but at a scale on the order of $\xi_T$ the flow crosses over toward the secondary classical critical fixed point, that controls it for large scales. In the presence of multiple dynamical scales, we find instead a quantum-to-classical crossover regime, which extends over a parametrically large range bounded by the two thermal lengths, $\xi^>_T \sim T^{-1/z_>}$ and $\xi^<_T \sim T^{-1/z_<}$. In this extended crossover regime, the transversal shear fluctuations ($z_< = 2$) still possess a quantum-mechanical character and drive the theory toward the primary quantum fixed point whereas the longitudinal mode ($z_> = 3$) is already classical and competes to push the theory toward the secondary classical fixed point, see also Fig.~\ref{fig:ModeMomenta}. 

We find that this extended quantum-to-classical crossover determines the critical properties in the quantum critical regime (III) of the phase diagram in Fig.~\ref{fig:PD}. Remarkably, the correlation length, $\xi$, depends here on temperature in a universal manner, Eq.~(\ref{UniversalXi}). This universality emerges via a peculiar interplay of logarithmic singularities of different origin. These singularities arise as the transversal quantum fluctuations have an effective dimension $d+z_< = d^+_c$ equal to the upper, $d^+_c = 4$, and the longitudinal classical fluctuations have an effective dimension $d = d^-_c$ equal to the lower critical dimension, $d^-_c = 2$, of the respective $\Phi^4$ theories. We showed that the interaction of these modes results in a $T$-dependent boost in the RG flow of the control parameter, Eq.~(\ref{accRGFlow}), that finally determines the $T$ dependence of $\xi$.

The two thermal lengths, $\xi^>_T$ and $\xi^<_T$, also identify crossover lines in the phase diagram Fig.~\ref{fig:PD} where the critical thermodynamics changes qualitatively. The sensitivity on these crossovers, however, differs for different thermodynamic quantities depending on whether they are dominated by the $z_>=3$ or the $z_<=2$ fluctuations. As a consequence, an interpretation in terms of scaling relations involving only a single dynamical exponent is, in general, not possible. 

To summarize, we found a series of phenomena arising from the presence of coexisting dynamics, which might be generic for quantum critical systems with multiple scales. So we identified (a) an extended quantum-to-classical crossover as summarized in Fig.~\ref{fig:ModeMomenta}, with an intermediate momentum regime where quantum and classical fluctuations coexist and their interaction profoundly affects the critical behavior. The multiple scales are (b) reflected in the crossover lines in the phase diagram, see Fig.~\ref{fig:PD}. We found that (c) certain type of thermodynamic quantities are governed by one or the other dynamics so that (d) simple scaling relations in terms of a single dynamical exponent are not applicable. Our study shows that multiple scales at quantum criticality can yield a rich phase diagram, complex critical properties, and even emerging universality.

\acknowledgments
We acknowledge insightful discussions and a collaboration with A. Rosch at an early stage of this work. 
We thank M. Vojta for useful discussions. This work was supported by the DFG through SFB 608 and through the Forschergruppe "Quantum phase transitions."

\appendix

\section{Effective fermion-boson theory for the Pomeranchuk quantum phase transition}
\label{app:FBModel}

In order to study the shear susceptibility close to the Pomeranchuk quantum critical point beyond the free-fermion form, viz., RPA, see Eqs.~(\ref{ShearSuscep2}), we consider the following fermion-boson model:
\begin{align} \label{FBModel}
\lefteqn{
\mathcal{S}[\Psi^\dag,\Psi,\sigma] = }
\\\nn&
\int d\tau d^d x \left[
\Psi^\dag g_0^{-1} \Psi  + \frac{1}{2} \sigma_{ij} \chi_{0ijkl}^{-1} \sigma_{kl} + \Gamma^{ijkl}_0 \sigma_{ij}\, \Psi^\dag \hat{Q}_{kl} \Psi \right],
\end{align}
where $g_0$ is the fermionic Green's function given below Eq.~(\ref{model}), $\chi_0$ is the shear susceptibility [Eq.~(\ref{ShearSuscep2})], and the vertex is $\Gamma_0^{ijkl} = \mathds{1}_{ijkl} \lambda$ with the coupling constant $\lambda$.

The bosonic self-energy in second order in $\lambda$, see Fig.~\ref{fig:diagrams}(a), will only shift the bare parameters of the susceptibilty $\chi_0({\bf q},i\Omega_n)$ but mantains the form of its $\Omega_n$ and $\bf q$ dependence. To indicate this renormalization, we will drop the index 0 and consider the parameters of $\chi$
as having effective values that have to be determined self-consistently. The corrections in higher order $\mathcal{O}(\lambda^4)$ to the quadrupolar polarization in Figs.~\ref{fig:diagrams}(b) and \ref{fig:diagrams}(c)  involve the lowest-order fermionic self-energy and vertex correction, respectively. We consider them first before turning to the discussion of the corrections to the shear susceptibility. 

\subsection{Fermion self-energy}
\label{sec:Self-Energy}

\begin{figure}
\includegraphics[width=0.4\textwidth]{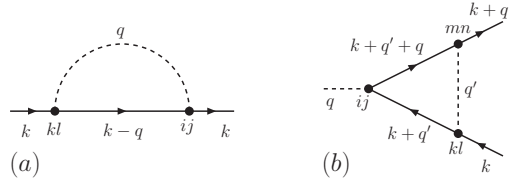} 
\caption{
(a) Fermion self-energy and (b) vertex correction of the model [Eq.~(\ref{FBModel})]. For clarity, we dropped the frequency labeling. 
}
\label{fig:Corrections}
\end{figure}

We calculate the fermionic  self-energy, $\Sigma = g^{-1} - g_0^{-1}$. In second order in the coupling $\lambda$, see Fig.~\ref{fig:Corrections}(a), it reads
\begin{align} \label{Sigma1}
\Sigma({\bf k},i\omega_n) &= - \lambda^2 \frac{1}{\beta} \sum_{{\bf q}, \Omega_n}  Q_{ij}({\bf k} - \frac{{\bf q}}{2}) Q_{kl}({\bf k} - \frac{{\bf q}}{2})
\nn\\&\times
\chi_{ijkl}({\bf q},i \Omega_n) g_0({\bf k-q},i\omega_n - i\Omega_n).
\end{align}
Taking into account that ${\bf q}$ is small compared to the Fermi wave vector this simplifies to
\begin{align} \label{Sigma2}
\Sigma({\bf k},i\omega_n) \approx& - \lambda^2  Q_{ij}({\bf k}) Q_{kl}({\bf k}) 
\\\nn\times&
\frac{1}{\beta} \sum_{{\bf q}, \Omega_n} \chi_{ijkl}({\bf q},i \Omega_n) \frac{-1}{i\omega_n - i \Omega_n - \varepsilon_k + {\bf v_F} {\bf q}}.
\end{align}
with ${\bf v_F}=v_F {\bf \hat{k}}$
The important contribution to the self-energy originates from the longitudinal part of the susceptibility
\begin{align}\label{Sigma3}
\Sigma({\bf k},i\omega_n) \approx&\, \lambda^2 \frac{1}{\beta} \sum_{\Omega_n} \int_0^\infty \frac{dq q}{(2\pi)^2}
 \chi_{\parallel\parallel}(q,i \Omega_n)
\\\nn& \times
\int_0^{2\pi} d\varphi \frac{2 \cos^2 2\varphi}{i\omega_n - i \Omega_n - \varepsilon_k + v_F q \cos\varphi}.
\end{align}
Close to the Fermi surface, this evaluates to
\begin{align}\label{Sigma4}
\Sigma(i\omega_n) &= - i \frac{\lambda^2}{\pi^2 v_F} {\rm sign}\, \omega_n
\int\limits_0^{|\omega_n|} d \Omega
\int\limits_0^\infty dq\, \chi_{\parallel\parallel}(q,i \Omega)
\\\nn
&= -\frac{\lambda^2}{v_F \xi_0} \left\{\!\begin{array}{ll}
\omega_0^{1/3}
i |\omega_n|^{2/3}  {\rm sign}\, \omega_n
& {\rm if}\; |\omega_n| \gg |r|^{3/2}
\\[1em]
\frac{1}{2\pi}  \frac{i \omega_n}{\sqrt{r}}
& {\rm if}\; |\omega_n| \ll r^{3/2}
\end{array}
\right.
\end{align}
where $\omega_0^{1/3} = \frac{1}{\pi \sqrt{3}} \left(\frac{v_F}{\eta \xi_0}\right)^{1/3}$. Note that the leading-order contribution only involves the frequency dependence. 

The self-energy of the form (\ref{Sigma4}) is well-known from the problem of electrons coupled to a Landau-damped gauge field\cite{Lee89,Altshuler94,Kim94} and ferromagnetic quantum criticality in metals.\cite{Rech06}

\subsection{Vertex correction}

The lowest order vertex correction, $\delta \Gamma^{ijkl} \!=\! \Gamma^{ijkl} \!-\! \Gamma^{ijkl}_0$, shown in Fig.~\ref{fig:Corrections}(b) reads
\begin{gather}
\delta\Gamma^{ijkl}_{{\bf k},i\omega_n}\!({\bf q},i\Omega_n) \!=\! 
\lambda^3 \frac{1}{\beta} \!\sum_{{\bf q'},\Omega'_n} 
Q_{ij}({\bf k}) Q_{mn}({\bf k})
\chi_{klmn}({\bf q'},i \Omega'_n)
\nn\\ 
\times \, {g_0}_{{\bf k+q'},i\omega_n + i\Omega'_n} \, {g_0}_{{\bf k+q+q'},i\omega_n + i \Omega_n + i \Omega'_n}
\end{gather}
where we used already that the bosonic momenta, ${\bf q}$ and ${\bf q'}$, are small compared to the fermionic momentum ${\bf k}$. Applying the algebraic identity $A B = - \frac{A-B}{A^{-1}-B^{-1}}$ to the product of fermionic Green's functions, the expression for the vertex correction can be rewritten as 
\begin{align} \label{VertexWard}
\delta\Gamma^{ijkl}_{{\bf k},i\omega_n}({\bf q},i\Omega_n) &\approx
\\\nn& \hspace*{-1em}
- \lambda \frac{F_{ijkl}({\bf k+q},i\omega_n+i \Omega_n)-F_{ijkl}({\bf k},i\omega_n)}{ i \Omega_n- {\bf v_F q}}.
\end{align}
Here, we introduced the auxiliary function
\begin{align}
F_{ijkl}({\bf k},i\omega_n) &= 
- \lambda^2 Q_{ij}({\bf k}) Q_{mn}({\bf k}) \\\nn
&\times \frac{1}{\beta} \sum_{{\bf q'},\Omega'_n}  \chi_{klmn}({\bf q'},i \Omega'_n)
{g_0}_{{\bf k+q'},i\omega_n + i\Omega'_n}.
\end{align}
Identifying $Q_{mn}({\bf k}) = \sqrt{2} E^\parallel_{mn}(\hat{\bf k})$ with the basis matrix defined in Eq.~(\ref{Eigenvectors}) and using the transformations [Eq.~(\ref{Trafos})] this simplifies to
\begin{align}
F_{ijkl}({\bf k},i\omega_n) =& - 2 \lambda^2 E^\parallel_{ij}(\hat{\bf k}) \frac{1}{\beta} \sum_{{\bf q'},\Omega'_n}  
{g_0}_{{\bf k+q'},i\omega_n + i\Omega'_n} 
\\\nn 
&\times \left(
E^\parallel_{kl}(\hat{\bf k}) \cos^2 2\phi \,\chi_{{\parallel \parallel} {\bf q'},i \Omega'_n} \right.\\\nn 
&\left.+ E^\perp_{kl}(\hat{\bf k}) \sin^2 2\phi \,\chi_{{\perp\perp} {\bf q'},i \Omega'_n} \right),
\end{align}
where $\hat{\bf k} \hat{\bf q}' = \cos\phi$.
In the low-energy limit, this expression reduces to $F_{ijkl}({\bf k},i\omega_n) \approx E^\parallel_{ij}(\hat{\bf k})E^\parallel_{kl}(\hat{\bf k}) \Sigma(i\omega_n)$ with the self-energy of Eq.~(\ref{Sigma4}). From Eq.~(\ref{VertexWard}) thus follows the asymptotic relation for the vertex correction
\begin{align} \label{VertexWard2}
\delta\Gamma^{ijkl}_{{\bf k},i\omega_n}({\bf q},i\Omega_n) \approx
- \lambda\, E^\parallel_{ij}(\hat{\bf k})E^\parallel_{kl}(\hat{\bf k})
\frac{ \Sigma_{i\omega_n+i \Omega_n}-\Sigma_{i\omega_n}}{ i \Omega_n- {\bf v_F q}}.
\end{align}
This relation can be identified as an asymptotic Ward identity\cite{Metzner98} deriving from fermion number conservation. 
The interaction vertex $\Gamma^{ijkl}_0$ in Eq.~(\ref{FBModel}) locally couples the fermion density $\Psi_{k+q}^\dag \Psi_k$ to the projection of the shear field onto the fermionic quadrupolar momentum tensor, i.e., $E^\parallel_{ij}(\hat{\bf k}) \sigma_{ij}({\bf q},i\Omega_n)$. (Note that $Q_{ij}({\bf k}) \approx \sqrt{2} E^\parallel_{ij}(\hat{\bf k})$.) This projection is thus a field conjugate to the density fluctuations. In other words, the projection of the vertex $\Gamma^{ijkl}_0 E^\parallel_{ij}(\hat{\bf k})E^\parallel_{kl}(\hat{\bf k})$ is just a local density-density interaction and obeys the corresponding Ward identity, from which follows Eq.~(\ref{VertexWard2}) in the low-energy limit.\cite{Metzner98} 

\subsection{Interaction corrections to the quadrupolar polarization}

We now consider the modification to the bosonic propagator $\chi$ in fourth order in the fermion-boson interaction $\lambda$ due to the diagrams (b) and (c) in Fig.~\ref{fig:diagrams}. In particular, we would like to check the stability of the dynamical part of $\chi$ and its associated dynamical exponents. The  strong longitudinal $z_>=3$ fluctuation are responsible for the anomalous fermionic self-energy [Eq.~(\ref{Sigma4})] of non-Fermi liquid form. This self-energy correction might feed back into the quadrupolar polarization of the electrons modifying the dynamics of the perpendicular mode.  So we focus in the following on the corrections to the perpendicular dynamics and show that its dynamical exponent $z_<=2$ is in fact robust against this feedback effect.

First, consider the contribution of diagram (b) to the perpendicular mode,  
\begin{align}
\delta \Pi_{\perp\perp}^{\rm (b)}({\bf q},i\Omega_n) \!=\!  \frac{4 i \nu \lambda^2}{(v_F q)^2} \int_0^{-\Omega_n} \!\!d\omega \left(\Sigma(i\omega+i\Omega_n) - \Sigma(i\omega)  \right).
\end{align}
Using the expression (\ref{Sigma4}) for the self-energy, at criticality, $r=0$, this reduces to 
\begin{align} \label{Result(b)}
\delta \Pi_{\perp\perp}^{\rm (b)}({\bf q},i\Omega_n) = - \frac{24}{5} \frac{\lambda^4 \nu }{v_F \xi_0} \omega_0^{1/3} \frac{|\Omega_n|^{5/3}}{(v_F q)^2}.
\end{align}
The resulting dynamics seems to dominate over the lowest order result for 
$\Pi_{\perp\perp}$ given in Eq.~(\ref{Pol0TransDyn}). Taking into account only diagram (b), one might naively expect a new dynamical exponent for the perpendicular mode given by $z_< = 12/5 > 2$. 

However, it turns out that the leading contribution of diagram (b) given in Eq.~(\ref{Result(b)}) is exactly compensated by the diagram (c) that includes the vertex correction,
\begin{align}
\delta \Pi_{ijkl}^{\rm (c)}({\bf q},i\Omega_n) &=
- \frac{\lambda}{\beta} \sum_{{\bf k}, \omega_n} 
\delta\Gamma^{ijmn}_{{\bf k},i\omega_n}({\bf q},i\Omega_{n})
\\\nn&\times Q_{mn}({\bf k}) Q_{kl}({\bf k})
{g_0}_{{\bf k+q},i\omega_n + i\Omega_n} {g_0}_{{\bf k},i\omega_n}
\end{align}
The cancelation of the leading contributions of diagrams (b) and (c) becomes apparent and is easily verified upon using the asymptotic identity for the vertex correction [Eq.~(\ref{VertexWard2})].

A similar cancellation of singular self-energy and vertex correction in the electron polarization due to the particle number conservation was discussed before in the context of the electron-gauge-field problem.\cite{Kim94} In particular, note that a Eliashberg-type of theory would disregard the diagram (c) in Fig.~\ref{fig:diagrams} containing the vertex correction and would erroneously conclude that the longitudinal $z_>=3$ mode couples back and modifies the dynamics of the perpendicular mode from $z_<=2$ to $z_<=12/5$.

\section{Angular averages over the shear mode propagator}
\label{sec:Averages}

For the computations of the perturbative corrections in Sec.~\ref{sec:RG} the following angular averages are needed
\begin{subequations}
\label{AngAverProd}
\begin{gather} 
\hspace*{-2em}\int \frac{d \hat{\bf q}}{2\pi} \,\chi^{\alpha \beta}_{{\bf q},i\Omega_n} = 
\frac{1}{2}\,
\left(
\chi_{\parallel\parallel{\bf q},i\Omega_n} +\chi_{\perp\perp{\bf q},i\Omega_n} 
\right)
\delta_{\alpha \beta},
\\ 
\hspace*{-2em}\int \frac{d \hat{\bf q}}{2\pi} \,\chi^{\alpha \beta}_{{\bf q},i\Omega_n} \chi^{\gamma \delta}_{{\bf q},i\Omega_n} 
=\qquad\qquad\qquad\qquad\qquad
\\\nn
\hspace*{-2.3em}\frac{1}{8}\left(\chi_{\parallel\parallel{\bf q},i\Omega_n} - \chi_{\perp\perp{\bf q},i\Omega_n}   \right)^2
\left(\delta_{\alpha \gamma} \delta_{\beta \delta} + \delta_{\alpha \delta} \delta_{\beta \gamma} 
\right)
\\\nn
\hspace*{.5em}+ \frac{1}{8} 
\left(
\chi^2_{\parallel\parallel{\bf q},i\Omega_n} \!+\! \chi^2_{\perp\perp{\bf q},i\Omega_n} \!+\! 6 \chi_{\parallel\parallel{\bf q},i\Omega_n} \chi_{\perp\perp{\bf q},i\Omega_n} \!
\right) \!\delta_{\alpha \beta} \delta_{\gamma \delta}.
\end{gather}
\end{subequations}

\section{Renormalization group at finite temperatures}
\label{app:RGT>0}

In this section, we extend the RG treatment of the effective model [Eq.~(\ref{EffShearTheory})] to finite temperatures in order to treat the quantum-to-classical crossover. Following Ref.~\onlinecite{Millis93}, we apply standard crossover theory\cite{Nelson75} to express the free energy as a line integral along an RG trajectory,
\begin{align} \label{FreeEnergy-RG}
\mathcal{F} = 
\int_0^\infty d \lambda\, e^{-(d+z)\lambda} 
\sum_{\alpha \in \{<,>\}}  f_\alpha \left(T(e^\lambda),r(e^\lambda),\eta_\alpha(e^\lambda)\right),
\end{align}
where $d=2$. 
We follow the considerations in Sec.~\ref{sec:RG} and leave the exponent $z$ unspecified in order to demonstrate 
the equivalence of 
different implementations of the RG procedure.
The running temperature, $T(b)$, control parameter, $r(b)$, and dynamical parameters, $\eta_\alpha(b)$ are defined below.  
The integration kernels are given by 
\begin{align}
 f_\alpha   &= \Lambda \frac{\partial}{\partial \Lambda} \frac{1}{2} 
\int \frac{d^2 q}{(2\pi)^2} \frac{1}{\beta}\sum_{\Omega_n} \log 
\left( \chi^{-1}_{\alpha\alpha}(q,i\Omega_n) \right)
\end{align}
where the identification $\perp \hat{=} <$ and  $\parallel \hat{=} >$ for the index $\alpha$ applies, and the susceptibilities are defined in Eqs.~(\ref{LongSuscep}) and (\ref{TransSuscep}).
The momentum integral and the sum over bosonic Matsubara frequencies is understood to be regularized with some UV cutoff $\Lambda$. Replacing the Matsubara sum with an integral by analytic continuation and employing a hard cutoff regularization, the function $f_\alpha$ becomes
\begin{align}
f_\alpha  &= \Lambda \frac{\partial}{\partial \Lambda} \frac{K_2}{2} 
\int_0^\Lambda dq q
\dashint_{-\Lambda^{z}}^{\Lambda^{z}} \frac{d \omega}{2\pi} 
\coth\left(\frac{\omega}{2T}\right)
\\\nn&\qquad\qquad\times
{\rm Im} \log \left(\chi^{-1}_{\alpha\alpha}(q,\omega+i 0)\right)
\end{align}
with $K_2 = 1/(2\pi)$ and a principal-value integral over frequencies $\omega$. Note that we choose to take the frequency cutoff to be $\Lambda^{z}$. It is convenient to separate the functions $f_\alpha$ into nonuniversal and universal parts, 
\begin{align}
f_\alpha(T,r,\eta_\alpha) = f_{0,\alpha}(r,\eta_\alpha) + f_{\infty,\alpha}(T,r,\eta_\alpha).
\end{align}
%
The nonuniversal parts, $f_{0,\alpha}$, are given by
\begin{align}
\label{NonUni1}
f_{0\,<}(r,\eta_<) &= f_<(0,0,\eta_<) + f^{(0,1,0)}_<(0,0,\eta_<) r,  \\
\label{NonUni2}
 f_{0\,>}(r,\eta_>) &= f_>(0,0,\eta_>) + f^{(0,1,0)}_>(0,0,\eta_>) r \\
 \nn& \quad + \frac{1}{2} f^{(0,2,0)}_>(0,0,\eta_>) r^2,
\end{align}
where $f_\alpha^{(0,n,0)}$ is the $n$th derivatives with respect to the second argument. They contribute considerably only at the initial stage of the RG flow, i.e., for small value of $\lambda$ in Eq.~(\ref{FreeEnergy-RG}), and give rise only to finite shifts of the bare values of the theory that can be absorbed by appropriate counterterms. The universal parts of the integration kernels, $f_{\infty,\alpha}$, read explicitly (for $\Lambda \to \infty$)
\begin{align} \label{IntKernelPerp}
f_{\infty,<}(T,r,\eta_<)  =& - \frac{1}{32 \pi \eta_<} r^2
\\\nn&
- \frac{\Lambda^2}{4\pi} \int_{\Lambda \sqrt{r+\Lambda^2}/\eta_<}^\infty d \omega \left[\coth\left(\frac{\omega}{2 T}\right)-1\right],
\\\label{IntKernelPara}
f_{\infty,>}(T,r,\eta_>) =\, & \frac{r^3 \Lambda}{15 \pi^2 \eta_> (r + \Lambda^2)}
\nn\\&
- \frac{\Lambda^2}{4 \pi^2} \int_0^\infty d \omega \left[\coth\left(\frac{\omega}{2 T}\right)-1\right] 
\nn\\& 
\times \arctan\left[\frac{\eta_> \omega}{\Lambda (r+\Lambda^2)}\right].
\end{align}
The first terms on the right-hand side give the $T=0$ contribution, respectively, and the second terms yield the finite temperature corrections. Both yield the universal critical contribution to the free energy. $\mathcal{F}_{\rm cr}$. Note that upon rescaling $r \to r \Lambda^2$, $T \to T \Lambda^z$ and $\eta_\alpha \to \eta_\alpha \Lambda^{z_\alpha - z}$, the dependence on the cutoff $\Lambda$ reduces to a global multiplication by $\Lambda^{d+z}$ with $d=2$, that can be later absorbed into the trivial scale dependence of the free energy. Furthermore, the $\eta$ dependence of the universal kernels reduces to 
\begin{align}
f_{\infty,\alpha}(T,r,\eta_\alpha) = \frac{1}{\eta_\alpha} f^{\rm red}_{\infty,\alpha}(\eta_\alpha T,r),
\end{align}
which simply follows after substitution $\omega \to \omega/\eta_\alpha$ in the integral of Eqs.~(\ref{IntKernelPerp}) and (\ref{IntKernelPara}). 

We distinguish between a longitudinal and transversal part of the free energy, $\mathcal{F}_{\rm cr} = \sum_{\alpha} \mathcal{F}_{{\rm cr}, \alpha}$, with
\begin{align} \label{crFreeEnergy-RG}
\mathcal{F}_{{\rm cr},\alpha} &= \int_0^\infty \!\! d \lambda\, e^{-(d+z)\lambda} \frac{1}{\eta_\alpha(e^\lambda)} 
f^{\rm red}_{\infty,\alpha}(\eta_\alpha(e^\lambda) T(e^\lambda),r(e^\lambda) ).
\end{align}
The running dynamical parameters $\eta_\alpha(b)$, temperature, $T(b)$, and control parameter, $r(b)$, entering Eq.~(\ref{crFreeEnergy-RG}) are given by the RG equations
\begin{align} \label{RGEqEta>T>0}
\frac{\partial \eta_>}{\partial \log b} & = (z_> - z) \eta_>,
\\\label{RGEqEta<T>0}
\frac{\partial \eta_<}{\partial \log b} & = (z_< - z) \eta_<,
\\
\frac{\partial T}{\partial \log b} &= z T,
\\
\label{RGEqMassT>0}
\frac{\partial r}{\partial \log b} &= 
2 r + \frac{2}{3} u \sum_{\alpha \in \{<,>\}} \frac{1}{\eta_\alpha} 
\frac{\partial}{\partial r} {f^{\rm red}_{\infty,\alpha}}(\eta_\alpha T, r),
%
%
\\\label{RGEqQuarticCoupT>0}
\frac{\partial u}{\partial \log b} &= (4- d - z) u - \frac{3}{32\pi \eta_<}  u^2,
\end{align}
where $d=2$, $z_< = 2$, and $z_> = 3$. The initial conditions are $\eta_\alpha(1) = \eta_\alpha$, $T(1) = T$, $r(1) = r$, and $u(1)=u$. 
The temperature dependence of the RG flow of the quartic coupling, $u$, can be neglected in the asymptotic low-energy limit so that Eq.~(\ref{RGEqQuarticCoupT>0}) coincides with Eq.~(\ref{RGQuarticCoup}).

One can convince oneself that, of course, the critical properties of physical observables do not depend on the specific form of frequency scaling employed in the RG process, i.e., on the value of the exponent $z$. In particular, note that the combinations of scaling variables $u(b)/\eta_\alpha(b)$ and $\eta_\alpha(b) T(b)$, are independent of $z$. As a consequence, the flow of the running control parameter $r(b)$ is also independent of $z$ as it is only determined by these combinations. Moreover, the additional multiplicative factor $1/\eta_\alpha(e^\lambda)$ in the integrand of Eq.~(\ref{crFreeEnergy-RG}) just cancels the dependence of the trivial scaling dimension of the free energy on $z$ so that $\mathcal{F}_{{\rm cr}}$ is also independent of the specific choice of $z$.

The correlation length is obtained from $r(b)$ with the help of the identification $\xi^{-2} = \left.r(b)/b^2\right|_{b=\Lambda \xi}$. The zero-temperature part of the transversal integration kernel, Eq.~(\ref{IntKernelPerp}), proportional to $r^2$ governs the $T=0$ flow of the control parameter. Only this leading term was kept in Eq.~(\ref{RGEqMass}). The neglected longitudinal kernel, Eq.~(\ref{IntKernelPara}), gives a subleading nonanalytic correction to $\xi^{-2}$ on the order $\mathcal{O}(u_{\rm eff} r^{3/2})$ at $T=0$ with the quartic coupling $u_{\rm eff}/\eta_< = u(\Lambda \xi)/\eta_<(\Lambda \xi)$ at the scale $b= \Lambda \xi$. On the other hand, the correlation length at finite temperatures in the regimes (II) and (III) of the phase diagram in Fig.~\ref{fig:PD} is controlled by the extended quantum-to-classical crossover. This crossover is associated with scales $b$ obeying the relation, $\eta_<(b/\Lambda) T(b/\Lambda) < 1 < \eta_>(b/\Lambda) T(b/\Lambda)$, which corresponds to an extended scaling range due to the presence of different dynamical exponents, $z_> \neq z_<$.
Solving for $\xi$ in these regimes, one recovers the self-consistent Eq.~(\ref{CorrelationLength}).

The critical thermodynamics presented in Sec.~\ref{sec:CrThermodynamics} follows from expression (\ref{crFreeEnergy-RG})
for the free energy. In particular, the critical compressibility in regimes (I) and (II), see Eq.~(\ref{CompLowT}), derives from the $T=0$ part of the transversal integration kernel [Eq.~(\ref{IntKernelPerp})].
\vspace*{1em}

\subsection{Scaling variables $\eta_>$ and $\eta_<$}
\label{app:Limits}

The RG Eqs.~(\ref{RGEqEta>T>0}) and (\ref{RGEqEta<T>0}) imply that either $1/\eta_>$ or $\eta_<$ is irrelevant when fixing the scaling dimension of the other dynamical parameter to zero, $z=z_<$ or $z=z_>$, respectively. Here, we would like to demonstrate that these parameters are in fact dangerously irrelevant. For this purpose, it is sufficient to show\cite{Cardy} that the critical thermodynamics is not well-defined in the two limits, $1/\eta_> \to 0$ and $\eta_< \to 0$.
First, consider the contribution to the critical free energy deriving from the perpendicular mode. At $T=0$, it simplifies to 
\begin{align} \label{Demo1}
\left.\mathcal{F}_{{\rm cr},<}\right|_{T=0} = - \frac{1}{32 \pi \eta_<} \int_0^\infty d \lambda\, e^{-4\lambda}   [r(e^\lambda)]^2.
\end{align}
From this expression follows the aforementioned critical compressibility in Eq.~(\ref{CompLowT}). As $\mathcal{F}_{{\rm cr},<}$ is proportional to the inverse of $\eta_<$, critical thermodynamics is obviously not well-defined in the limit $\eta_< \to 0$. Second, consider $\mathcal{F}_{{\rm cr},>}$ deriving from the longitudinal mode at criticality $r=0$. The leading temperature dependence is given by
\begin{align} \label{Demo2}
\left.\mathcal{F}_{{\rm cr},>}\right|_{r=0, T\to 0} &=
\frac{1}{\eta_>} \int_0^\infty d \lambda\, e^{-5 \lambda} 
f^{\rm red}_{\infty,>}(\eta_> T e^{3 \lambda} ,0 ) 
\nn\\&
\sim \frac{1}{\eta_>} (\eta_> T)^{5/3}.
\end{align}
The dependence on $\eta_>$ and $T$ obtains after the appropriate substitution $\lambda \to \lambda - \frac{1}{3} \log (\eta_> T)$ in the integral and taking the limit of infinite cutoff, $\Lambda \to \infty$, which eliminates the residual $T$ dependence due to the lower boundary of the integration range. From Eq.~(\ref{Demo2}) follows the leading behavior of the specific heat in regimes (II) and (III), Eq.~(\ref{SpecHeat-III}). Clearly, 
it follows from Eq.~(\ref{Demo2}) that thermodynamics is also ill-defined in the limit $\eta_> \to \infty$. Taken together, this identifies $\eta_<$ and $1/\eta_>$ as dangerously irrelevant variables for the respective choices of $z$.


\begin{thebibliography}{99}


\bibitem{Sachdev}
S.~Sachdev, {\it Quantum Phase Transitions}
(Cambridge University Press, Cambridge, England, 1999).

\bibitem{ZinnJustin}
J.~Zinn-Justin, {\it Quantum Field Theory and Critical Phenomena} (Cambridge University Press, Cambridge, England, 2002).

\bibitem{Hertz76}
J.~A. Hertz, Phys. Rev. B {\bf 14},  1165  (1976).

\bibitem{Moriya95} 
T.~Moriya, {\em Spin Fluctuations
in Itinerant Electron Magnetism} (Springer-Verlag, Berlin, 1985).
T.~Moriya and T.~Takimoto, J. Phys. Soc. Jpn. {\bf 64}, 960
(1995).

\bibitem{Millis93}
A.~J. Millis, Phys. Rev. B {\bf 48},  7183  (1993).

\bibitem{Loehneysen07}
H. v. L\"ohneysen, A. Rosch, M. Vojta, and P. W\"olfle,
Rev. Mod. Phys. {\bf 79}, 1015 (2007).

\bibitem{Gegenwart08}
P.~Gegenwart, Q.~Si, and F. Steglich, Nat. Phys. {\bf 4}, 186 (2008).

\bibitem{Gegenwart07}
P. Gegenwart, T. Westerkamp, C. Krellner, Y. Tokiwa, S. Paschen, C. Geibel, F. Steglich, E. Abrahams, and Q. Si,
Science {\bf 315}, 969 (2007).


\bibitem{Vojta97}
T. Vojta, D. Belitz, R. Narayanan, and T. R. Kirkpatrick, Z. Phys. B: Condens. Matter {\bf 103}, 451 (1997).

\bibitem{Belitz00}
D. Belitz, T. R. Kirkpatrick, R. Narayanan, and T. Vojta, Phys. Rev. Lett. {\bf 85}, 4602 (2000).


\bibitem{Belitz04}
D. Belitz, T. R. Kirkpatrick, and J. Rollb\"uhler, Phys. Rev. Lett. {\bf 93}, 155701 (2004).

\bibitem{Rech06}
J. Rech, C. P\'epin, and A. V. Chubukov,
Phys. Rev. B {\bf 74}, 195126 (2006)

\bibitem{Abanov04}
A. Abanov and A. V. Chubukov, Phys. Rev. Lett. {\bf 93}, 255702 (2004).


\bibitem{Pomeranchuk54}
I. Pomeranchuk, Sov. Phys. JETP {\bf 8}, 361 (1958).

\bibitem{Oganesyan01}
V. Oganesyan, S. A. Kivelson, and E. Fradkin, Phys. Rev. B {\bf 64}, 195109 (2001).

\bibitem{Metzner03}
W. Metzner, D. Rohe, and S. Andergassen, Phys. Rev. Lett. {\bf 91}, 066402 (2003).


\bibitem{DellAnna06}
L.~Dell'Anna and W.~Metzner, Phys. Rev. B {\bf 73}, 045127 (2006).


 
\bibitem{Yamase04}
H. Yamase, Phys. Rev. Lett. {\bf 93}, 266404 (2004).

\bibitem{Yamase05}
H. Yamase, V. Oganesyan, and W. Metzner, Phys. Rev. B {\bf 72}, 035114 (2005).

\bibitem{Yamase07}
H. Yamase, Phys. Rev. B {\bf 76}, 155117 (2007).
 
\bibitem{Yamase09}
H. Yamase, Phys. Rev. Lett. {\bf 102}, 116404 (2009).


\bibitem{Lawler06}
M. J. Lawler, D. G. Barci, V. Fernandez, E. Fradkin, and L. Oxman, Phys. Rev. B {\bf 73}, 085101 (2006).

\bibitem{Chubukov06}
A. V. Chubukov and D. V. Khveshchenko, Phys. Rev. Lett. {\bf 97}, 226403 (2006).

\bibitem{Lawler07}
M. J. Lawler and E. Fradkin, Phys. Rev. B {\bf 75}, 033304 (2007).


\bibitem{Woelfle07}
P. W\"olfle and A. Rosch, J. Low Temp. Phys. {\bf 147}, 165 (2007).

\bibitem{Quintanilla06}
J. Quintanilla and A. J. Schofield, Phys. Rev. B {\bf 74}, 115126 (2006).


\bibitem{Quintanilla08}
J. Quintanilla, M. Haque, and A. J. Schofield, Phys. Rev. B {\bf 78}, 035131 (2008).

\bibitem{Ho08}
A. F. Ho and A. J. Schofield, EPL {\bf 84}, 27007 (2008).

\bibitem{Vojta09}
M. Vojta, arXiv:0901.3145

\bibitem{Borzi07}
R. A. Borzi, S. A. Grigera, J. Farrell, R. S. Perry, S. J. S. Lister, S. L. Lee, D. A. Tennant, Y. Maeno, and A. P. Mackenzie, Science {\bf 315}, 214 (2007).

\bibitem{Fradkin07}
E. Fradkin, S. A. Kivelson, and V. Oganesyan, Science {\bf 315}, 196 (2007).
 

\bibitem{Sachdev97}
S. Sachdev, Phys. Rev. B {\bf 55}, 142 (1997).


\bibitem{Lee89}
P. A. Lee, Phys. Rev. Lett. {\bf 63}, 680 (1989).

\bibitem{Altshuler94}
B. L. Altshuler, L. B. Ioffe, and A. J. Millis, Phys. Rev. B {\bf 50}, 14048 (1994).

\bibitem{Kim94}
Y.~B.~Kim, A.~Furusaki, X.-G.~Wen, and P.~A.~Lee, Phys. Rev. B {\bf 50}, 17917 (1994).

\bibitem{Metzner98}
W. Metzner, C. Castellani, and C. Di Castro,
Adv. Phys. {\bf 47}, 317 (1998).

\bibitem{Cardy}
John L. Cardy, {\it Scaling and Renormalization in Statistical Physics} (Cambridge University Press, Cambridge, England, 1996).


\bibitem{Zhu03}
L.~Zhu,  M.~Garst, A.~Rosch, and Q.~Si,
Phys. Rev. Lett. {\bf 91}, 066404 (2003).

\bibitem{Anfuso08}
F. Anfuso, M. Garst, A. Rosch, O. Heyer, T. Lorenz, C. R\"{u}egg, and K. Kr\"amer,
Phys. Rev. B {\bf 77}, 235113 (2008).

\bibitem{Nelson75}
D. R. Nelson, Phys. Rev. B {\bf 11}, 3504 (1975).

\end{thebibliography}
\end{document}